\documentclass[graybox,envcountchap]{svmult}
\usepackage[square, sort&compress, numbers]{natbib}
\usepackage[pdfpagelabels=true, plainpages=false,
            colorlinks=true, allcolors=blue]{hyperref}

\usepackage{url}
\usepackage{mathptmx}       % selects Times Roman as basic font
\usepackage{helvet}         % selects Helvetica as sans-serif font
\usepackage{courier}        % selects Courier as typewriter font
\usepackage{type1cm}        % activate if the above 3 fonts are
\usepackage{makeidx}         % allows index generation
\usepackage{graphicx}        % standard LaTeX graphics tool
\usepackage{multicol}        % used for the two-column index
\usepackage[bottom]{footmisc}% places footnotes at page bottom
\usepackage{amsmath}
\newcommand{\sign}{\text{sign}}

\makeindex             % used for the subject index
\begin{document}

\title*{Mutual synchronization in spin torque and spin Hall nano-oscillators}
\titlerunning{Mutual synchronization in spin torque and spin Hall nano-oscillators} 

\author{Akash Kumar, Artem Litvinenko, Nilamani Behera, Ahmad A. Awad, Roman Khymyn, and Johan~\AA kerman}
\authorrunning{Akash Kumar \textit{et al.}} 
% your contribution title if the original one is too long
\institute{Akash Kumar, Ahmad A. Awad and Johan \AA kerman* \at Department of Physics, University of Gothenburg, 41296, Gothenburg, Sweden, and also with Research Institute of Electrical Communication (RIEC) and Center for Science and Innovation in Spintronics (CSIS), Tohoku University, 2-1-1 Katahira, Aoba-ku, Sendai 980-8577 Japan.
\\(*e-mail: johan.akerman@physics.gu.se)\\
Artem Litvinenko, Nilamani Behera and Roman Khymyn \at Department of Physics, University of Gothenburg, 41296, Gothenburg, Sweden}
%
% Use the package "url.sty" to avoid
% problems with special characters
% used in your e-mail or web address
%
\maketitle

\abstract{%Spintronic oscillators can be classified into spin-torque and spin Hall nano-oscillators based on the mechanism of spin current generation \textit{i. e.} spin transfer torque effect and spin Hall effect, respectively. The synchronization of these nanoscopic spintronic oscillators, both to an external signal or mutual, leads to higher output power and lower operational linewidth along with the possibility of non-linear interaction between them. Mutual synchronization has been demonstrated as a potential candidate for unconventional computing. We first discuss the basic concept of spintronic oscillators and the possible mechanisms for synchronization. We then discuss the experimental demonstration of mutual synchronization in spin-torque and spin Hall nano-oscillators. Later the potential of mutual synchronization for wireless communication and unconventional computing is briefly discussed. Finally, we outline the future direction of mutual synchronization of these oscillators and their possible potential for various applications. 
This chapter reviews the state of the art in mutually synchronized spin-torque and spin Hall nano-oscillator (STNO and SHNO) arrays. After briefly introducing the underlying physics, we discuss different nano-oscillator implementations and their functional properties concerning frequency range, output power, phase noise, and modulation rates. We then introduce the concepts and the theory of mutual synchronization and discuss the possible coupling mechanisms in spintronic nano-oscillators, such as dipolar, electrical, and spin-wave coupling. We review the experimental literature on mutually synchronized STNOs and SHNOs in one- and two-dimensional arrays and discuss ways to increase the number of mutually synchronized nano-oscillators. Finally, the potential for applications ranging from microwave signal sources/detectors and ultrafast spectrum analyzers to neuromorphic computing elements and Ising machines is discussed together with the specific electronic circuitry that has been designed so far to harness this potential.\\
\\
This is a preprint of the following chapter: Akash Kumar, Artem Litvinenko, Nilamani Behera, Ahmad A. Awad, Roman Khymyn, and Johan~\AA kerman, \textit{Mutual synchronization in spin torque and spin Hall nano-oscillators}, published in Nanomagnets as Dynamical Systems, edited by Supriyo Bandyopadhyay,
Anjan Barman, 2024, Springer, reproduced with permission of the Springer Nature Switzerland AG 2024. The final authenticated version is available at: \href{https:
//doi.org/10.1007/978-3-031-73191-4}{https:
//doi.org/10.1007/978-3-031-73191-4}.
}

\tableofcontents

\section{Introduction}

The theoretical prediction of spin transfer torque (STT) by Slonczewski~\cite{Slonczewski1996a} and Berger~\cite{berger1996emission} in 1996 shaped spintronics beyond giant magneto-resistance (GMR)~\cite{Baibich1988,binasch1989enhanced} and tunneling magneto-resistance (TMR)~\cite{Miyazaki1995,Moodera1995}. According to STT, a spin-polarized current can transfer angular momentum to the local magnetization of a ferromagnetic thin film. The STT hence allowed manipulation of the magnetization dynamics using spin-polarized charge currents and led to the demonstration of both memory and oscillator devices known as spin-transfer torque magnetic random access memory (STT-MRAM)~\cite{Katine2000,Akerman2005sci} and spin-torque nano-oscillators (STNOs)~\cite{Chen2016procieee}, respectively. The first STNOs~\cite{Tsoi1998} were based on all-metallic nano-contacts and the GMR effect. Later, researchers realized nanopillar-based STNOs using both all-metallic structures and insulating spacer based magnetic tunnel junctions (MTJ), where the much higher tunneling magnetoresistance results in much stronger microwave output power~\cite{Kiselev2003,Chen2016procieee}. Their nanoscopic size and wide frequency tunability make these oscillators potential candidates for various applications including wireless communication, unconventional computing, and energy harvesting. However, their relatively low output power and broad spectral linewidth limit their commercial utilization. 

In the last decade, much emphasis has been placed on spintronic devices based on the Spin Hall effect (SHE), a phenomenon first predicted by Dyakonov and Perel~\cite{Dyakonov1971jetp}, and later revisited by Hirsch~\cite{Hirsch1999}. The spin Hall effect allows the generation of pure spin currents in non-magnetic materials due to their large spin-orbit coupling. In 2012, Liu \emph{et al.,}~\cite{Liu2012prl}, demonstrated, for the first time, auto-oscillation using the giant spin Hall effect in $\beta$-Ta thin films~\cite{liu2012science}. However, this demonstration still included an MTJ as an integral part of the read-out. 
In independent work, Demidov \emph{et al.,}~\cite{demidov2012nm}, showed that SHE-based spintronic oscillators do not require an MTJ pillar and can generate auto-oscillations in simple non-magnet/ferromagnet bilayers that could be electrically read using the anisotropic magnetoresistance (AMR). These oscillators are known as spin Hall nano-oscillators (SHNOs). Though SHNOs result in low output power compared to STNOs, the ease of fabrication and direct optical access provide a large playground to study different properties, including mutual synchronization.

Since the first STNO demonstrations, researchers have been looking into mutual synchronization of these devices~\cite{Kaka2005,Mancoff2005,Houshang2015natnano,tsunegi2018scaling}, as it can be used to both increase the microwave output power and reduce its linewidth. In the last two decades, various methods of synchronization have been proposed and demonstrated, which led to improved spectral quality and rich interactions between oscillators that can be leveraged for emerging unconventional computing applications. In this chapter, we will discuss the mutual synchronization in both STNOs and SHNOs. In the following sections, we will first discuss the working principle and architecture of these devices and various mechanisms for synchronization to external signals and each other. Later, we will benchmark these synchronized oscillators for their output power and spectral parameters, followed by a discussion of various applications that can benefit from mutual synchronization.

\section{Spintronic oscillators}

\subsection{Mechanism of operation}

\begin{figure}[b!]
    \centering
    \includegraphics[width=\linewidth]{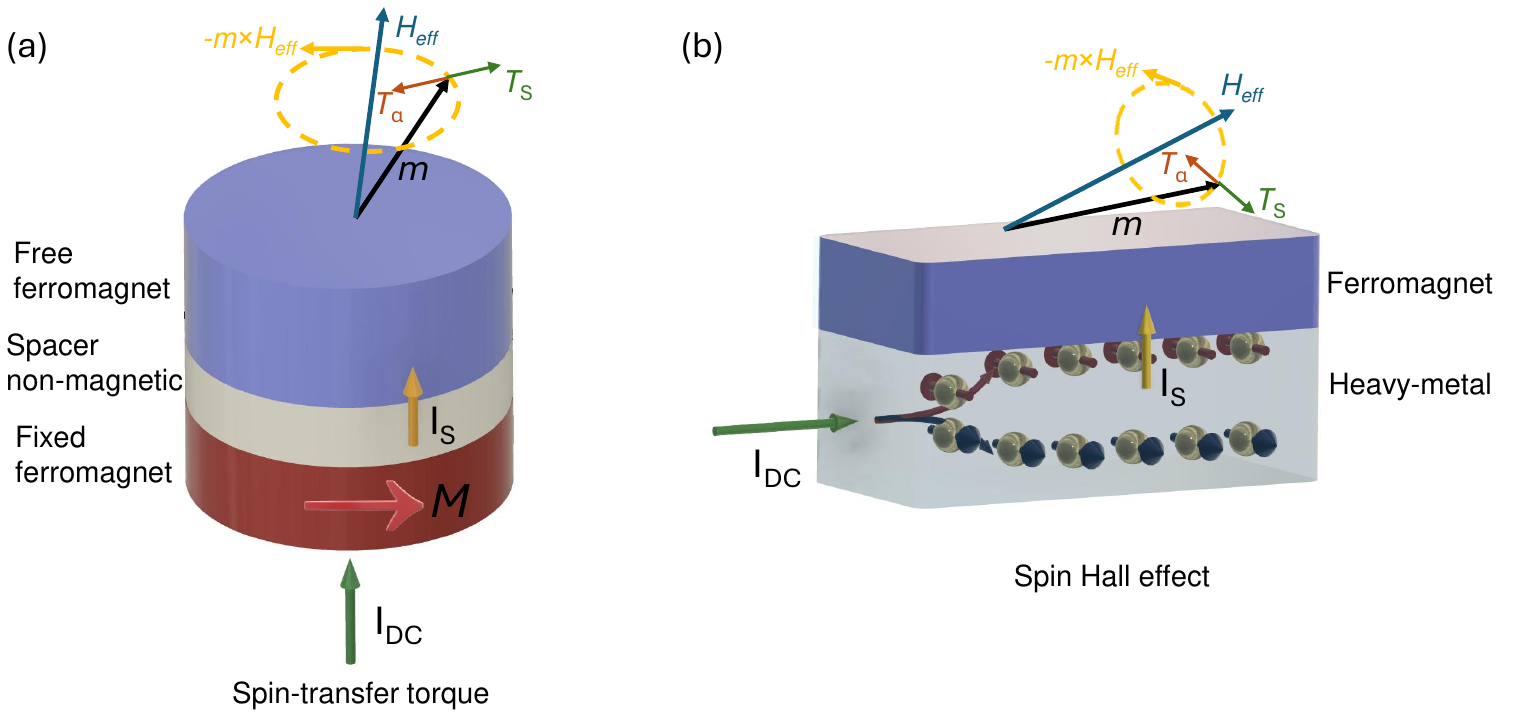}
    \caption{Schematic representation of (a) spin-transfer torque (STT) between two ferromagnetic (FM) layers, and (b) the spin Hall effect (SHE) from a heavy metal layer. The spin-polarized current generated by STT, and the pure spin current generated by the SHE, exert an anti-damping torque on the FM layer and can, when the local damping is cancelled, drive the FM into auto-oscillation.}
    \label{fig:STT_SHE}
\end{figure}

Figure~\ref{fig:STT_SHE} shows the schematic representation of the basic principle for the operation of STNOs and SHNOs, where $M$ is a fixed magnetic layer, whose dynamics is often ignored (\emph{e.g.}~due to greater thickness, higher damping, or being pinned to an adjacent antiferromagnet), and $m$ is a free layer where all the large angle magnetodynamics is excited. As discussed earlier, STNOs work on the principle of STT (shown in Fig.~\ref{fig:STT_SHE}a) mediated by spin-polarized currents between non-collinear ferromagnetic layers~\cite{Slonczewski1996a,ralph2008jmmm}. Figure~\ref{fig:STT_SHE}b shows a schematic for the spin Hall effect (SHE), where a charge current flowing through the heavy metal generates a pure spin current (free of net charge) in the transverse direction~\cite{Hirsch1999,sinova2015revmodern}. This intriguing effect arises from the significant spin-orbit coupling present in these materials. The resulting pure spin current accumulates at the interface between the non-magnetic and ferromagnetic materials, eventually diffusing into the ferromagnetic layer, where it exerts a torque similar to STT, also known as spin-orbit torque (SOT). 

The magnetization dynamics in the free ferromagnetic layer in the presence of %external 
magnetic fields %($\textbf H_\mathit{{eff}}$)
and STT/SHE is governed by the Landau-Lifshitz-Gilbert-Slonczewski (LLGS) equation:

\begin{equation} \label{eq:LLGS}
\frac{d \textbf m}{d t} = \gamma[\textbf H_\mathit{{eff}} \times \textbf m] + \textbf T_{\alpha} + \textbf T_{S}.
\end{equation}

The first term on the right-hand side represents the uniform conservative precession of the magnetization around the effective field ($\textbf H_\mathit{{eff}}$), which includes % H_\mathit{{eff}}$ consists of 
the applied external magnetic field $\textbf H_{a}$, the magneto-dipolar field $ \textbf H_\mathit{{dip}}$, the anisotropy field $\textbf H_{k}$, the exchange field $\textbf H_{ex}$, and, when a drive current is present, the Oersted field $\textbf H_{oe}$~\cite{slavin2009nonlinear,dvornik2018pra}:
\begin{equation} \label{eq:H_eff}
\textbf H_\mathit{{eff}}= \textbf H_{a}+ \textbf H_\mathit{{dip}}+\textbf H_{k}+\textbf H_{ex}+\textbf H_{oe}.
\end{equation}

The second term ($\textbf T_{\alpha}$) describes the (positive) damping term added by Gilbert~\cite{Gilbert1955}. The third term ($\textbf T_{S}$) comprises the positive or negative damping effects from the STT/SOT phenomena. For simplicity, we have assumed that the STT/SOT effects are purely of a damping-like character in Eq.1. Including their field-like components simply adds an extra STT/SOT-dependent term in $\textbf H_\mathit{{eff}}$.

At a material and field-dependent threshold current density (with certain current polarity), the anti-damping torque from the STT/SOT generated spin current density can be large enough to completely cancel the intrinsic damping in the system, which, under certain conditions, results in sustained oscillations of the magnetization. This is the operating principle of STNOs and SHNOs. Under different conditions, a sufficiently large STT/SOT can instead switch the magnetization, in a precessional manner, between different stable static directions. This is the operating principle of STT/SOT-MRAM. The LLGS equation can be mapped onto a general auto-oscillating model to understand and study the behavior of spintronic oscillators, described in detail by Slavin and Tiberkevich in their Tutorial~\cite{slavin2009nonlinear}.

Besides the frequency of magnetic precession, a non-uniform $\textbf H_\mathit{{eff}}$ will define the localization of the oscillatory mode, by creating a spatial potential profile for magnons. Also, it determines the strength and type of spin wave scattering, i.e. repulsion and attraction of spin waves. In the first case, the oscillation frequency increases with amplitude (positive non-linearity), while in the second, it decreases (negative non-linearity).

STNOs and SHNOs are usually made of thin magnetic films, and, hence, the shape anisotropy with the corresponding demagnetizing field $ \textbf H_\mathit{{dip}}$ aligns the magnetisation in the film plane. The demagnetization is reduced at the edges of the constriction, which creates edge spin wave modes that are especially relevant for SHNOs~\cite{dvornik2018pra}. The in-plane magnetization is characterized by a negative non-linearity, i.e., attraction between magnons. Hence, the excitation of solitons, such as spin wave bullets, is possible in this configuration~\cite{dvornik2018pra,rajabali2023injection}. The shape anisotropy can be counteracted by a perpendicular magnetic anisotropy (PMA), which arises at the interface between magnetic and adjunct layers. Often, a MgO top layer is employed to reach high values of $\textbf H_{k}$. Thus, by varying $\textbf H_{k}$ it is possible to create effectively isotropic samples~\cite{sethi2023compensation} or even to pull the magnetisation completely perpendicular to the film plane, when $|\textbf H_{k}|>|\textbf H_{dip}|$. The latter case is characterized by a strong attraction between magnons and, particularly, used to excite spin wave droplets~\cite{fulara2019spin}. 

The externally applied magnetic field $\textbf H_{a}$ is easy to change in the experiment, and can be used not only to tune the frequency, but also to change the auto-oscillating spin wave mode. For example, a perpendicular or sufficiently oblique $\textbf H_{a}$ pulls the magnetisation out of the film plane, and changes the nonlinearity from negative at near in-plane configuration to positive at high out-of-plane angles~\cite{bonetti2010experimental}. The latter case is favorable for the excitation of propagating spin waves. 

The exchange field $\textbf H_{ex}$ manifests itself in spatially inhomogeneous spin wave modes. Particularly it defines the profiles of bullets and droplets as well as the wave-vector of propagating spin waves at a given frequency. The Oersted field $\textbf H_{oe}$, created by an electric current flowing through the sample, usually has lower values in comparison to the other fields, but can still manifest itself in distinct ways, due to its different symmetry. For example, it adds a circular in-plane field component in STNOs \cite{Dumas2013prl,jiang2018impact} breaks the symmetry between the two constriction edges in SHNOs~\cite{dvornik2018pra}. Thus, the two SHNO edge modes, typically have different threshold currents, frequencies, amplitudes, and mode volumes.

\subsection{Device Architecture}

\begin{figure}
    \centering
    \includegraphics[width=\linewidth]{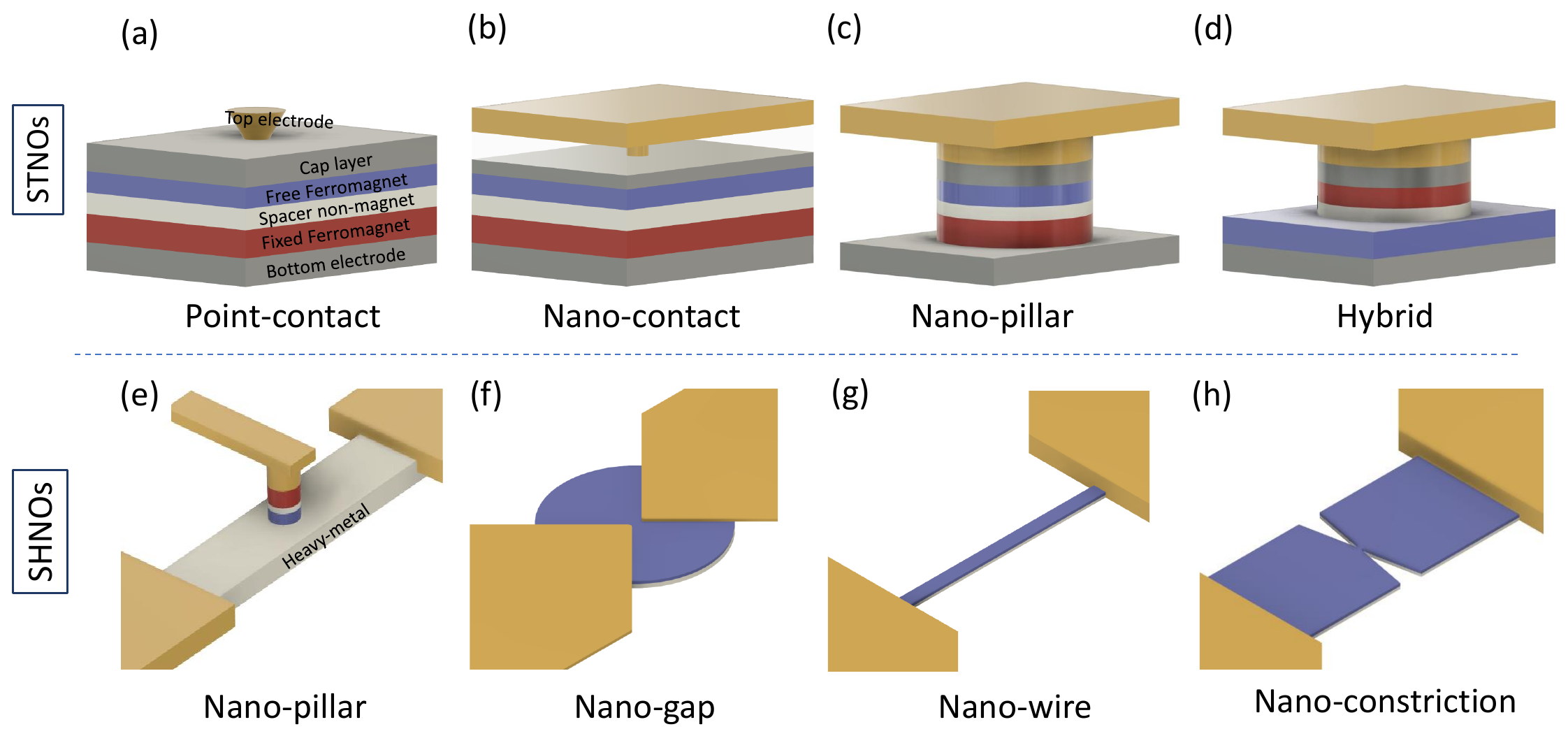}
    \caption{Various device geometries. STNOs: (a) point-contact, (b) nano-contact, (c) nano-pillar, and (d) hybrid. SHNOs: (e) nano-pillar, (f) nano-gap, (g) nano-wire, and (h) nano-constriction. }
    \label{fig:Osc_Architecture}
\end{figure}

Over the last two decades, various device architectures have been proposed and demonstrated for both STNOs and SHNOs, each with their unique properties and applicability. Figure~\ref{fig:Osc_Architecture} briefly summarizes the different device geometries. For more details on the operation and microwave signal properties of many of these devices, the reader may refer to a recent comprehensive review article~\cite{Chen2016procieee}. In its simplest form, an STNO consists of a thin film heterostructure, either based on GMR or TMR (MTJ) material stacks. The auto-oscillation amplitude in these devices can be electrically measured via the resistance oscillations generated by the GMR or TMR ratio, which under constant current generates a measurable microwave voltage. Figure~\ref{fig:Osc_Architecture}a shows the initial attempt of point-contact STNOs fabricated as a mechanical point contact using a highly sharpened metallic tip~\cite{Tsoi1998,Tsoi2000,Sun1999}. With later advances in lithography techniques, nano-contacts were instead fabricated using e-beam lithography (shown in Fig.~\ref{fig:Osc_Architecture}b)~\cite{Myers1999,Rippard2003,Pufall2003}. In both the point-contact and nano-contact geometries, a large current density is applied through the nanoscopic contact on top of the STNO stacks. %The electrons flow from the bottom electrode generates spin polarized electron (due to spin dependent scattering from fixed ferromagnet). 
The large spin-polarized current exerts anti-damping torque on the free layer to generate auto-oscillations. Since the free layer is extended, it can support a range of different auto-oscillation modes, such as localized spin wave bullets~\cite{slavin2005prl,bonetti2010experimental,bonetti2012prb}, propagating spin waves~\cite{madami2011natnano,bonetti2012prb,madami2015prb}, magnetic droplets~\cite{mohseni2013science,chung2016natcomm,chung2018prl,ahlberg2022natcomm}, dynamical skyrmions~\cite{zhou2015natcomm}, and vortex gyration modes~\cite{pribiag2007magnetic}. The field-free operation has been demonstrated using free layers with perpendicular magnetic anisotropy~\cite{mohseni2011pssrrl} and exchange-spring free layers~\cite{jiang2023nl}. However, the extended nature of the material stack leads to very rapid lateral current spreading underneath the nano-contact, which results in both relatively large required current densities of the order of 10$^8$--10$^9$ A.cm$^{-2}$~\cite{Katine2000,Kiselev2003} and reduced microwave power. The significantly improved performance was achieved when the bottom electrode was made substantially thicker and, therefore, more conducting to mitigate the initial lateral current spreading in the GMR portion of the material stack~\cite{banuazizi2017order}. 
The lateral current spread in the nano-contact geometry also precludes the use of MTJ stacks, as the current will mostly pass through the metallic free layer instead of through the tunneling barrier. However, by etching the free layer into a tapered, "sombrero" shape~\cite{Maehara2013large,maehara2014high,Houshang2018natcomm}, one can still force a large part of the STNO current through the tunneling barrier and make use of a large fraction of the total TMR.

Figure~\ref{fig:Osc_Architecture}c shows the nano-pillar geometry typically used for STT-MRAM unit cells. Thanks to their fully patterned and truly nanoscopic size, the required current density is much lower ($<$ 10$^7$ A.cm$^{-2}$) compared to nano-contacts. %point/nano-contact STNOs ($>$10$^8$-10$^9$  A.cm$^{-2}$)~\cite{Katine2000,Kiselev2003}. 
As their cylindrical geometry precludes any propagating spin waves, the auto-oscillation modes are typically divided into uniform modes~\cite{Kiselev2003,litvinenko2022ultrafast} and vortex mode~\cite{pribiag2007magnetic,Martin2013}. Contrary to uniformly magnetized STNOs, the vortex based STNOs have a magnetic vortex (a strongly non-uniform magnetic texture) in the free ferromagnetic layer. In a magnetic vortex, the in-plane magnetization twists around the vortex core, which has a perpendicular out-of-plane orientation. Close to the core, the circular in-plane magnetization gradually leaves the plane and acquires an out-of-plane component as it orients itself towards the perpendicular direction of the core~\cite{zvezdin2022spin, ekomasov2021effect}. When a spin-polarized current flows through the pillar, the magnetic vortex core gets displaced from its equilibrium state at the center of the pillar and can enter various dynamic modes including gyroscopic motion~\cite{Jenkins2016}. As the current is completely confined to the pillar, these geometries are optimal for MTJ-based devices, which for vortex gyroscopic modes can yield record high microwave power~\cite{tsunegi2016microwave} and record low linewidths. Figure~\ref{fig:Osc_Architecture}d shows a hybrid nano-pillar geometry that uses the nano-pillar structure for the spin polarization layers but the free layer as an extended thin film, which allows for propagation of spin waves and also a desired direct optical access of the free layer~\cite{Demidov2010,Evarts2013}.  

For SHNOs, a simple ferromagnet/non-magnet bilayer is sufficient, and the oscillations can be electrically detected using AMR. Figure~\ref{fig:Osc_Architecture}e illustrates the initial demonstration of self-sustained auto-oscillations driven by the SHE in a nano-pillar geometry consisting of an MTJ pillar over a wider channel of SHE material ($\beta$-Ta)~\cite{Liu2012prl,cai2023angular}. Subsequently, nano-gap SHNOs were introduced~\cite{demidov2012nm}, as depicted in Fig.~\ref{fig:Osc_Architecture}f. These devices feature two pointed contact pads made of highly conductive gold (Au) with nano-gaps ranging from 100 to 500 nm, positioned over a nano-disk of FM/NM~\cite{demidov2012nm,Ranjbar2014,spicer2018time}. The charge current flows through the Au and generates a large current density in the nano-gap region. Figure~\ref{fig:Osc_Architecture}g shows the nano-wire SHNOs, first demonstrated by Duan \textit{et al}.~\cite{Duan2014}, where the FM/NM layers are patterned into nano-wires to generate the required current densities. Lastly, Fig.~\ref{fig:Osc_Architecture}h shows the schematic representation for nano-constriction SHNOs, where the FM/NM layers are patterned into a bow-tie-shaped nano-constriction~\cite{Demidov2014}. The nano-constriction not only helps in generating the required current density for auto-oscillations but also gives an over all lower resistance compared to the nano-wire and better heat conduction. Compared to STNOs, the latter two geometries are much easier to fabricate (using only a single high precision lithography step) and provide direct optical access to the magneto-dynamical modes. Another prominent SHNO geometry is domain wall-based SHNOs~\cite{sato2019domain}, which are similar to nano-wire SHNOs.

Over the last decade, nano-constriction SHNOs have been optimized using different material combinations in the FM/HM heterostructures, such as in-plane system of Py/Pt~\cite{durrenfeld2017nanoscale}, Py/W~\cite{Mazraati2016apl}, Py/AuPt~\cite{xie2023nanoscale}, bipolar Py/Pt/CoFeB~\cite{hache2020bipolar}, moderately perpendicular W/CoFeB/MgO~\cite{zahedinejad2018cmos,fulara2019spin,kumar2022fabrication}, W-Ta/CoFeB/MgO~\cite{behera2022energy} and perpendicularly magnetized Ta/Pt/[Co/Ni]/Co~\cite{choi2022voltage,sethi2023compensation}. Single FM layer SHNOs have also been demonstrated~\cite{haidar2019natcomm}. Different magnetization orientations (in-plane, moderately perpendicular to fully perpendicular) allow the generation of different spin-waves modes~\cite{dvornik2018pra,fulara2019spin}. Moreover, the constriction width also changes the non-linearity of the auto-oscillations~\cite{awad2020apl}. To date, the best spin Hall efficiency is found using W-Ta alloys~\cite{behera2022energy}, which allows auto-oscillations in a 10 nm SHNO with a record low current of 28 $\mu$A~\cite{behera2022energy}. Most of these demonstrations are based on highly resistive Si substrates, which also makes these oscillators CMOS compatible~\cite{zahedinejad2018cmos}.

In the remaining chapter, we will mainly focus on nano-contact and nano-pillar STNOs and nano-constriction SHNOs due to their established superiority in experimentally demonstrated mutual synchronization.

\section{Synchronization of spintronic oscillators}
The synchronization of both STNOs and SHNOs has been experimentally demonstrated with both external signals (injection locking) and with each other (mutual). Here, we will discuss the mechanism governing various synchronization/coupling schemes.

\subsection{Forced Synchronization: Locking to an external signal} 
Injection locking is a fundamental non-linear phenomenon in which an external signal enforces its frequency onto an otherwise free-running non-linear auto-oscillator. Due to their inherent magnetic nature, spintronic oscillators~\cite{Chen2016procieee} typically have an unusually strong non-linearity, and can hence be readily injection locked with large locking bandwidths~\cite{rajabali2023injection}, either via a microwave current~\cite{Rippard2005, zhou2008apl,Georges2008,romera2016apl,jue2018apl,Tortarolo2018scirep,hem2019prb,letang2019prb,Hache2019APL} or a microwave field\cite{urazhdin2010fractional, zhou2007jap, singh2017integer}. As discussed in more detail in Section 8, below, injection locking of such oscillators has recently been used for  various applications, such as phase shift keying\cite{litvinenko2021analog}, and neuromorphic computing.\cite{romera2018nt,zahedinejad2020nt,mcgoldrick2022ising,finocchio2023roadmap,chumak2022advances} STNOs and SHNOs can also be locked to its own frequency using a feedback circuit, which can substantially reduce the generation linewidth. The injection locking also allows synchronization of auto-oscillatory signal to sub-harmonic and fractional harmonic external signals~\cite{urazhdin2010fractional,singh2017integer}. Depending on the amplitude and frequency (same or multiple/fractional) of the driving signal, the locking can behave differently, ranging from bi-stable signals to chaotic behaviors. Synchronization to an external signal is also highly useful to study the phase of oscillators, \emph{e.g.}~using phase-resolved Brillouin light scattering (BLS) microscopy and for other fundamental properties.

\begin{figure}
    \centering
    \includegraphics[width=\linewidth]{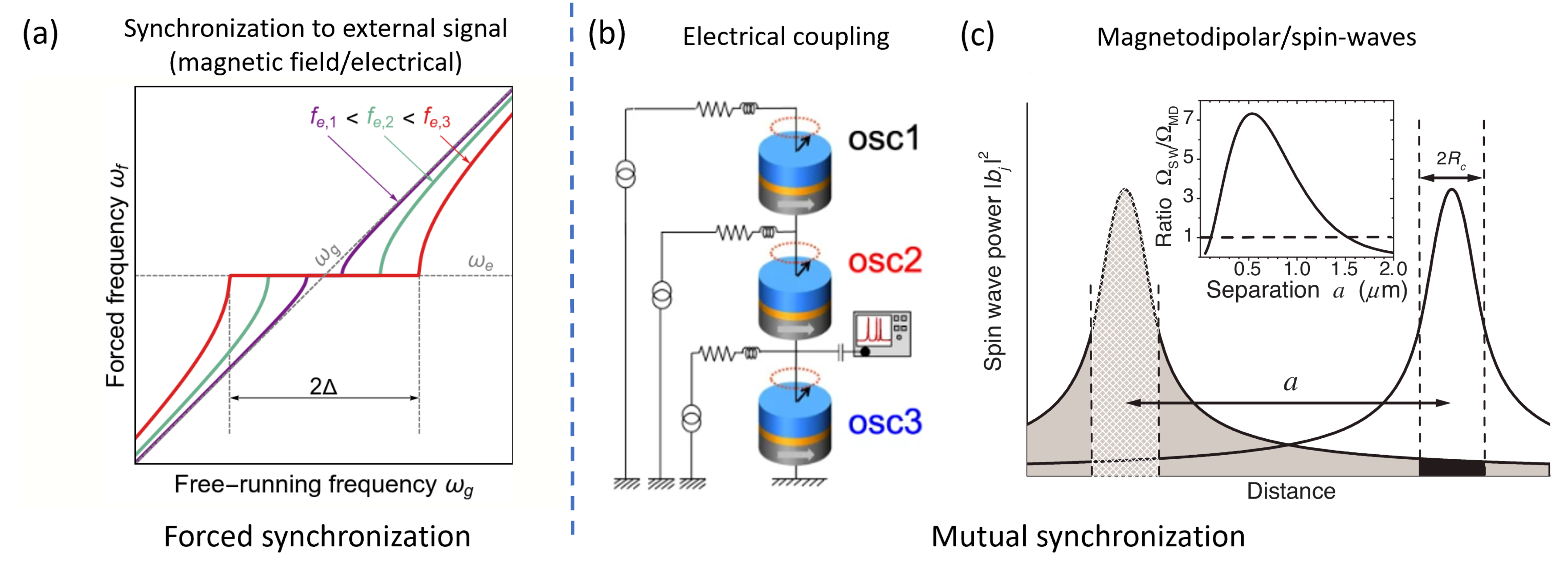}
    \caption{ (a) Forced synchronization, or injection locking, \textit{i.e.}~synchronization to an external signal, either electrical or from a microwave magnetic field, the data is similar to ~\cite{rajabali2023injection}. (b) Schematic illustration of electrical synchronization driven by RF currents generated by each oscillator. Reprinted from~\cite{romera2018nt}. (c) Distance dependence of magneto-dipolar and spin-wave driven mutual synchronization in STNOs. Reprinted with permission from ~\cite{Slavin2006}. }
    \label{fig:Coupling}
\end{figure}

\subsubsection{Injection locking theory}

The basic theory, which describes the synchronization phenomena of STNOs, was developed in \cite{slavin2005nonlinear,slavin2009nonlinear} by transforming Eq. \ref{eq:LLGS} to the known equation of a nonlinear oscillator:
\begin{equation}
    \frac{dc}{dt}+i\omega(|c|^2)c+\Gamma_+(|c|^2)c-\Gamma_-(|c|^2)c=\xi(t),
    \label{eq:oscillator}
\end{equation}
where $c(t)$ is a complex amplitude of the precession, $\xi(t)$ is an external signal acting on the oscillator, $\omega$ the nonlinear (power dependent) precession frequency, $\Gamma_+$ describes energy losses, usually mainly caused  by Gilbert damping, and $\Gamma_-$ the anti-damping action of STT/SOT. Note that all oscillator parameters depend on the oscillation power $|c|^2$, expressing the highly nonlinear nature of the oscillator. The first terms of the power expansion 
\begin{align}
\omega= & \omega_0+N|c|^2\\
\Gamma_+= & \Gamma_G(1+Q|c|^2)\\
\Gamma_-= & \sigma I(1-|c|^2)
\label{eq:expansion}
\end{align}
contain parameters of the nonlinear frequency shift $N$ and nonlinear damping $Q$. $I$ is the applied current, $\sigma$ describes the STT efficiency, $\omega_0$ is the frequency of the local FMR, and in $\Gamma_G=\alpha \omega_0$, $\alpha$ is the Gilbert damping constant. The nonlinear parameters define stationary values of the oscillation frequency $\omega_g$ and power $p_0$. It is convenient to introduce the dimensionless nonlinear parameter $\nu=N/(Q\Gamma_G+\sigma I)$ for further analysis. Eq. \ref{eq:oscillator} is widely used for the description of the dynamics of STNOs and SHNOs and their arrays. However, it should be used with caution for SHNOs since it does not take into account the variation of the mode profile and the volume of the oscillation mode with amplitude, which is quite common in unrestricted planar geometries.

\begin{figure}[t!]
    \centering
    \includegraphics[width=0.9\linewidth]{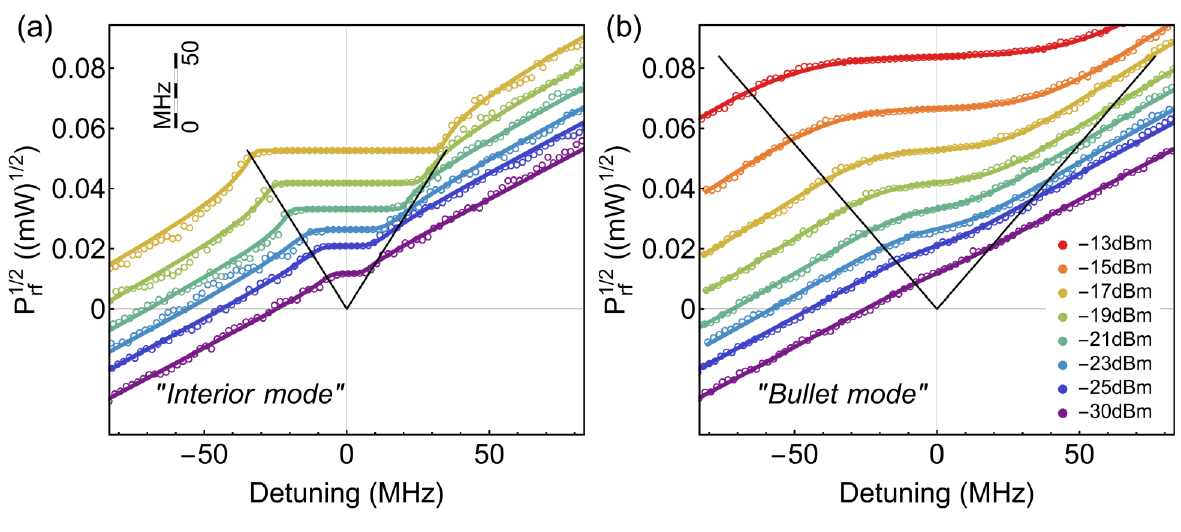}
    \caption{Injection locking of (a) a linear-like mode, and (b) a spin-wave bullet mode, in an SHNO, as a function of detuning. Open circles show the experimental data, while solid color lines -- analytical calculations by Eq. \ref{eq:detuningWithNoise}. Solid black lines show the dependence of the locking bandwidth on the amplitude of the driving signal, similar to Eq.\ref{eq:lockingBW}. Reprinted from~\cite{rajabali2023injection}.}
    \label{fig:InjectionLocking}
\end{figure}

Injection locking is described by Eq. \ref{eq:oscillator} with the driving term $\xi(t)=f_e \exp(-i\omega_e)t$, where $f_e$ and $\omega_e$ are the complex amplitude and frequency of the injected signal. In the locked regime the oscillation frequency is equal to $\omega_e$ within a certain detuning range $|\omega_e-\omega_g|<\Delta$, known as the locking bandwidth. For the model (Eqs. \ref{eq:oscillator}-\ref{eq:expansion}):

\begin{equation}
 \Delta=|f_e|\sqrt{(1+\nu^2)/p_0}.
 \label{eq:lockingBW}
\end{equation}
As the nonlinear parameter is quite large for STNOs and SHNOs ($\nu>1$) and has a primary contribution to the synchronization properties. 

Outside of the phase locking range, an oscillator experiences 'frequency pulling', where the forced frequency $\omega_f$ increasingly deviates from the external frequency and gradually approaches its intrinsic auto-oscillation frequency (see Fig. \ref{fig:Coupling}(a)):
\begin{equation}
    \omega_f=\omega_e+\sign(\omega_g-\omega_e)\sqrt{(\omega_g-\omega_e)^2-\Delta^2}.
\end{equation}

\subsubsection{The impact of noise}

The synchronization properties of STNOs and SHNOs deteriorate substantially in the presence of noise, due to their inherently weak total oscillation power and small mode volume. It is possible to analytically account for impact of noise in most cases, assuming low power of the noise and the RF injection, since this allows for reduction of Eq.~\ref{eq:oscillator} to the Adler-like equation for the oscillator's phase ~\cite{rajabali2023injection}:
\begin{equation}
\frac{d \psi}{dt}=\Delta_f- \Delta \sin{\psi}+\frac{\xi(t)}{\sqrt{p_0}},
\label{eqAdler}
\end{equation}
where $\psi(t)=\phi(t)-\omega_et$ is the difference between the phase of the oscillator $\phi(t)$ and RF signal with an angular velocity $\omega_{e}$ and effective amplitude $\epsilon_{e} \sim p_{e}^{1/2}$, $\Delta$ is the one-sided locking BW in the absence of the noise defined in Eq. \ref{eq:lockingBW}, and $\xi(t)$ describes $\delta$-like correlated noise $\langle \xi \xi_{\tau}\rangle= k p_0 \delta(\tau)$, where $k$ defines the dispersion of the noise, \emph{i.e.}~the HWHM generation linewidth of the free running oscillator.

Equation \ref{eqAdler} describes the motion of the oscillator phase as a Brownian particle in the periodical inclined potential $V(\psi)=-\Delta_f \psi - \Delta\cos\psi$, where the detuning $\Delta_f$ defines the potential inclination.
Thermal, white Gaussian, noise is unlimited in amplitude and inevitably leads to phase slips between potential minima with the differrent probabilities of slips in opposite directions, due to the inclination. Employing a Fokker-Planck formalism, one can define the averaged dynamics of phase slips and find a final result for the oscillator forced frequency as  \cite{rajabali2023injection}:
\begin{equation}
\omega_f-\omega_e=k\sinh\frac{\Delta_f}{k}|I_{2i\Delta_f/k}(2\Delta/k)|^{-2},
\label{eq:detuningWithNoise}
\end{equation}
where $I_{in}(x)$ denotes a Bessel function of the pure  imaginary order $n$.

According to Eq. \ref{eq:detuningWithNoise}, the presence of noise ’softens’ the edges of the locking bandwidth, more so for modes with a low ratio of auto-oscillation to noise power. This can lead to substantially different locking behavior for modes with different volumes even if the locking bandwidths are very similar. Particularly, the impact of noise on the injection locking was studied for spin wave bullets \cite{rajabali2023injection, Demidov2014b} and linear-like modes \cite{rajabali2023injection}. The direct comparison reveals a substantially higher impact of noise on the bullet mode due to its smaller volume and, hence, smaller total power ~\cite{rajabali2023injection} (Fig. \ref{fig:InjectionLocking}).
%for example, for the soliton, the bullet mode (small volume) and the linear-like one (larger volume) in SHNOs~\cite{rajabali2023injection, Demidov2014b}, see Fig. \ref{fig:InjectionLocking}.

\subsubsection{Fractional injection locking}
It has been demonstrated by several groups \cite{urazhdin2010fractional,singh2017integer,Lebrun2015} that the STNO can be locked to the external microwave signal not only at an integer multiple of its auto-oscillation frequency but also at
fractional multiples above and below the fundamental. Fractional synchronization can be utilized to obtain better phase noise characteristics and avoid overlapping of external signal harmonics and subharmonics with a synchronized nano-oscillator.

\subsection{Mutual Synchronization}
Due to their low power and large linewidth, compared to their CMOS counterparts, spintronic oscillators have not yet seen any commercial success. Mutual synchronization of these oscillators holds the highest promise for their commercial applications, not just for wireless communications but also for emergent computing applications (see Section 8). Mutual synchronization can be governed by various coupling mechanisms, such as magneto-dipolar coupling, current-driven electrical coupling, and coupling via spin waves. Dipolar coupling %due to the oersted magnetic field need 
requires that the oscillators be fabricated close to each other, as it rapidly decays with distance. Electrical synchronization needs a large electrical signal, which so far has only been achieved using MTJs. % to lock the oscillator frequencies. 
Coupling via propagating spin waves first requires a common magnetic layer, and second, propagating spin waves can be excited, which is not always the case but can be realized, by using perpendicular magnetic anisotropy and/or a large applied field. Different oscillator geometries (discussed in Sec. II) hence behave differently and have varying dominant coupling mechanisms. For example, nano-contact STNOs share their ferromagnetic free layers and in strong perpendicular (or tilted) fields are mainly synchronized using spin-wave coupling. MTJ nano-pillars can be fabricated sufficiently close to each other for strong dipole interactions. The large output signal from nano-pillar MTJs is also useful for electrical synchronization. Nano-constriction SHNOs have the possibility to utilize all coupling mechanisms thanks to their easy nearby fabrication and shared free layers. However, for sufficient electrical coupling, MTJs will have to be integrated with them.   

A simple theoretical analysis of mutual synchronization begins with a set of coupled Eqs.~\ref{eq:oscillator}, with the right-hand side accounting for the signals from neighboring oscillators \cite{slavin2009nonlinear}:
\begin{equation}
    \xi_j(t)=\Omega_{j,j'}\exp({i\beta_{j,j'})c_{j'}(t)},
\end{equation}
where $\Omega$ defines the coupling strengths and $\beta$ the coupling phase (or delay). For the simplest case of two coupled oscillators, mutual phase-locking is possible within the frequency difference range:
\begin{equation}
|\omega_{g,1}-\omega_{g,2}|<\Delta_2, \qquad \Delta_2=2\Omega\sqrt{1+\nu^2}|\cos(\beta-\arctan\nu)|.
\label{eq:mutualBW}
\end{equation}
Thus, the nonlinearity significantly impacts mutual synchronization not only by increasing the maximum frequency bandwidth but also by setting the condition for the coupling phase $\beta$. The maximum locking bandwidth is achieved for
\begin{equation}
\beta=\arctan\nu +n\pi,
\label{eq:couplingPhase}
\end{equation}
when $n$ is an integer and vanishes for half-integer $n$. 

In the phase locked regime both oscillators generate auto-oscillations at the same common frequency:
\begin{equation}
\omega=\left(\omega_{g,1}+\omega_{g,2}\right)/2-\sqrt{\Delta_2^2-\left(\omega_{g,1}-\omega_{g,2}\right)^2}\tan(\beta-\arctan\nu),
\end{equation}
and the relative phase between oscillators depends on the detuning of their free-running frequencies as $\phi_1-\phi_2=\arcsin\left((\omega_{g,1}-\omega_{g,2}\right)/\Delta_2)$ if $\cos(\beta-\arctan\nu)>0$ or $\phi_1-\phi_2=\pi-\arcsin\left((\omega_{g,1}-\omega_{g,2}\right)/\Delta_2)$ otherwise.

\subsubsection{Spin wave coupling} 
If the coupling between oscillators is provided by propagating spin waves (SWs), the coupling phase increases linearly with distance $a$ and SW wave vector $k$ as $\beta=k a$.
The locking bandwidth is hence an alternating function of the inter-oscillator spacing, according to the condition of Eq.~\ref{eq:mutualBW}, and even closely placed oscillators can hence fail to synchronize at particular frequencies. The phase difference between the oscillators in the synchronized regime varies between 0 and $\pi$ with distance, i.e. between in-phase and anti-phase synchronization \cite{Kendziorczyk2014,Kendziorczyk2016prb}.  

Coupling via propagating SWs usually dominates within the practical separation range below 1 $\mu$m~\cite{Kaka2005, Pufall2006} (see, Fig. \ref{fig:Coupling}(c)). The coupling strength decays with oscillator separation due to the radial emission of SWs, and natural damping as $\Omega\propto\sqrt{r_c/a}\exp(-\Gamma_G a/v)$, where $r_c$ is the radius of the nano-contact and $v$ is the group velocity of the propagating SWs. The first decay mechanism can be greatly reduced by SW channelling into narrow beams, using either a patterning of the magnetic film or the local magnetic field configuration. In particular, it was demonstrated in \cite{Houshang2015natnano} that the Oersted field landscape results in a highly anisotropic SW propagation in STNOs based on nanocontact, which promotes synchronization between two oscillators observed for separations larger than 1 $\mu$m. The usage of SWs with a high group velocity can further expand the range of synchronization. 

\subsubsection{Magneto-dipolar coupling} 
Magneto-dipolar coupling is an intrinsic feature affecting all STNOs and SHNOs regardless of the presence of other coupling mechanisms. In contrast to other mechanisms, it does not require a shared magnetic layer or a shared current. The dipolar coupling is conservative, $\beta$ is real and independent of distance, making it appealing for synchronizing nano-pillar STNOs, including vortices \cite{Erokhin2014, flovik2016describing}. The coupling strength $\Omega$ decreases rapidly with increasing oscillator separation as $1/a^3$, \emph{i.e.}, in the same way as the magnetic field of a point magnetic dipole. In systems with a shared magnetic free layer, the dipolar coupling, therefore, dominates over SWs at very short distances ($<100$nm). Unless the SWs are formed into beams, dipolar coupling also dominates at large distances ($>1\mu$m) but is too weak for any mutual synchronization to occur in practice \cite{slavin2009nonlinear} (see, Fig. \ref{fig:Coupling}(c)).

\subsubsection{Direct exchange} 
Direct exchange between precessing spins in two different auto-oscillating regions may also provide coupling. This was \emph{e.g.}~argued in Ref. \cite{awad2017natphys}, where long-range mutual synchronization between two nano-constriction SHNOs separated by up to 4 µm was reported as due to the tails of the auto-oscillating regions extending deep into the bridges and producing coupling. At the same time, these extending regions will also produce dipolar coupling, and a detailed study separating the importance of the two mechanisms has yet to be carried out. Mutual synchronization of closely spaced vortices underneath four nano-contacts fabricated on a uniform magnetic film has also been ascribed to direct exchange between spins in the form of antivortices \cite{RuotoloA.2009}.

\subsubsection{Electrical coupling} 
Electrical coupling requires a spintronic oscillator with sufficient output power to be connected electrically in series or parallel~\cite{romera2018nt,tsunegi2018scaling,sharma2021electrically}.  Electrical coupling is convenient for oscillators without shared free magnetic layers, such as nano-pillars. The output of one oscillator can be fed to the inputs of its neighbors through an external circuit, where the signal can be amplified, attenuated or shifted in phase, which provides a dynamic configurability of the oscillator network. The free-running frequencies of each oscillator can be tuned individually by controlling the current at each oscillator, further extending the flexibility of the operation. Recently, a network of four electrically coupled STNOs was trained to recognize spoken vowels by tuning the frequencies according to the real-time learning rule~\cite{romera2018nt} (see Section 8, below).

\section{Mutual synchronization of spin torque nano-oscillators}

In the following section, we will review different STNO designs and discuss their corresponding preferable mutual synchronization mechanisms:

\begin{figure}[b!]
    \centering
    \includegraphics[width=\linewidth]{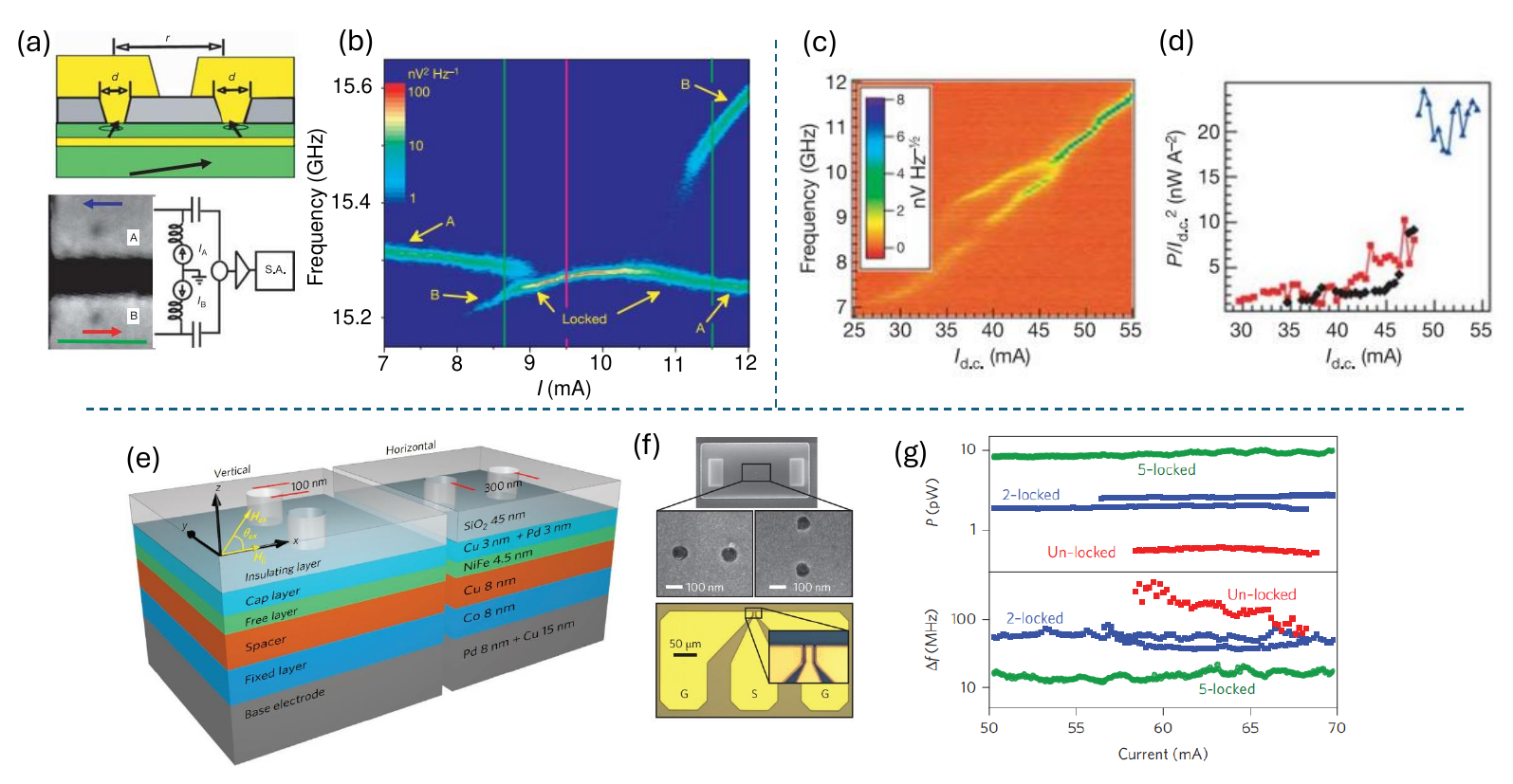}
    \caption{ (a) Schematic illustrating the cross-sectional view of two nano-contact STNOs with diameter, \textit{d} and center to center separation, \textit{r}. The bottom image shows the scanning micro-graph along with the measurement setup. The STNOs are connected to two individual current sources ($I_A$ and $I_B$). (b) Power spectral density (PSD) from both of the nano-contacts with varying current, $I_B$ in nano-contact B (with current in nano-contact A fixed to, $I_A$ = 8.0 mA). (c) PSD versus current ($I_{dc}$) for two nano-contacts connected in parallel with a single top contact. (d) Integrated output power normalized with the square of the charge current; the synchronized mode at high current shows significantly larger output power. (e) Schematic of an all-metallic spin valve with horizontal and vertical arrangement of STNOs with respect to the applied magnetic field. (f) Device geometry and scanning electron micrograph. (g) Output power and linewidth for unlocked, 2-locked, and 5-locked STNOs. The Fig. clearly demonstrates that mutual synchronization leads to higher output power and lower linewidth. Figures reprinted with permission from ~\cite{Kaka2005,Mancoff2005,Houshang2015natnano}.  }
    \label{fig:Nano_Contact}
\end{figure}

\subsection{Nano-contact STNOs}
Nano-contact STNOs with a continuous magnetic stack under the metallic nano-contact intrinsically allows for %the best conditions 
mutual synchronization via propagating spin waves~\cite{Kaka2005,Mancoff2005,Sani2013ntc,Houshang2015natnano}. In Fig.~\ref{fig:Nano_Contact}, we present different designs and their corresponding synchronization properties. In 2005, Kaka \textit{et al.}~\cite{Kaka2005} and Mancoff et al.~\cite{Mancoff2005}, independently presented the first synchronization studies on nano-contact STNOs. Figure~\ref{fig:Nano_Contact}a shows the geometry used by Kaka \textit{et al.}~\cite{Kaka2005}, where two STNOs are connected to different current sources (I$_A$ and I$_B$), and their high-frequency power outputs are combined and measured using a spectrum analyzer. Figure~\ref{fig:Nano_Contact}b shows the PSD \emph{vs.}~I$_B$ (current in STNO 'B'), with I$_A$ = 8.0 mA. It is remarkable that mutual synchronization could be achieved at a distance of 500 nm. Figure~\ref{fig:Nano_Contact}c $\&$ d shows the experimental results reported by Mancoff \textit{et al.}, where two STNOs share both top and bottom electrodes. As demonstrated in \cite{pogoryelov2011b,pogoryelov2012combined}, mutual synchronization by propagating spin waves is strong enough so that the pair of nano-contact STNOs stay locked even under strong current modulation, which confirms the robustness of mutual synchronization and its prospect for large STNO arrays. Later, in~\cite{Houshang2015natnano} it was demonstrated that up to 5 nano-contact STNOs can be successfully synchronized when the SWs are formed into directional beams. The results are summarized in Fig.~\ref{fig:Nano_Contact}e-g.

\subsection{Nano-pillar STNOs}
The design of nano-pillar STNOs is technologically more complex but allows for localized vortex gyration modes that demonstrate high stability, high power, and low phase noise~\cite{Kiselev2003}. Nano-pillars can also sustain more uniform modes but these suffer from worse signal properties, most likely since they involve the lower-quality perimeter in the auto-oscillation, which the vortex mode largely avoids. Mutual synchronization of adjacent nano-pillar STNOs can be achieved by dipolar interaction of closely spaced STNOs or direct electrical coupling between STNOs arbitrarily far apart. 

\begin{figure}[b!]
    \centering
    \includegraphics[width=\linewidth]{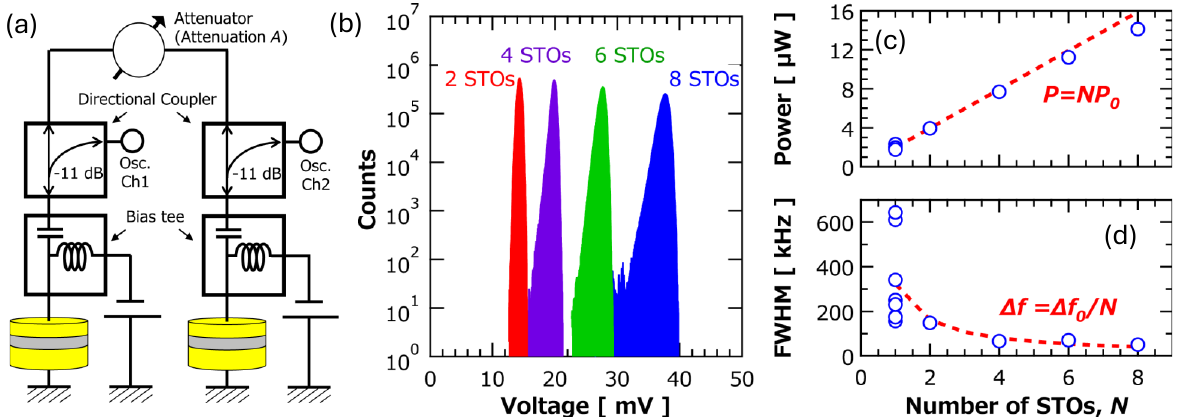}
    \caption{ (a) The measurement schematic used for mutual synchronization of STNOs. It is important to note that in this study the authors used individual magnets for each oscillator to manipulate their frequency into the locking range. Mutual synchronization is achieved by means of electrical coupling. (b) Histograms depicting the envelope of the RF voltage acquired from an array of multiple STNOs. (c) Maximum output power and (d) minimum linewidth (FWHM) versus number of oscillators (\textit{N}) connected in series. The dotted line represents the fit to $NP_0$ and $\Delta f/N$. Reprinted from ~\cite{tsunegi2018scaling} }
    \label{fig:tsunegi8STNOs}
\end{figure}

In order to exploit dipolar interaction between pillar the spacing between them should be less than 200 nm for pillars of 50 nm diameter~\cite{castro2022mutual}. This requirement is difficult to implement technically since the technology of nano-pillar STNO requires multi-angle etching of the magnetic stack. During the multi-angle magnetic stack etching, the neighboring pillars shade themselves, leading to the effect of magnetic stack bridges between the pillars, which consequently breaks the oscillation conditions. Nevertheless, mutual synchronization of dipolarly coupled vortex-based STNO nano-pillars separated by 100 nm was demonstrated in \cite{Locatelli2015}. It was shown that the dipolar interaction is more efficient for the synchronization of STNOs with interacting vortices gyrating in opposite directions. While magneto-dipolar coupling is hard to control,  it was shown that the synchronization properties could be tuned by varying a current in an additional intermediate oscillator \cite{abreu2016controlling}. 

Thanks to the high output power of MTJ nano-pillar STNOs, they can also be mutually synchronized with direct electrical connection \cite{lebrun2017mutual,tsunegi2018scaling}. In \cite{lebrun2017mutual} it was demonstrated that with electrical connection, the linewidth of two synchronized STNOs improves by a factor of 2 and, consequently, the phase noise improves by 3 dB. Tsunegi \textit{et al.}~\cite{tsunegi2018scaling} have scaled the mutual synchronization to up to 8 STNOs connected electrically. Figure~\ref{fig:tsunegi8STNOs}a-c summarizes the experimental results.

\subsubsection{Series and parallel synchronization} 

\begin{figure}[h!]
    \centering
    \includegraphics[width=\linewidth]{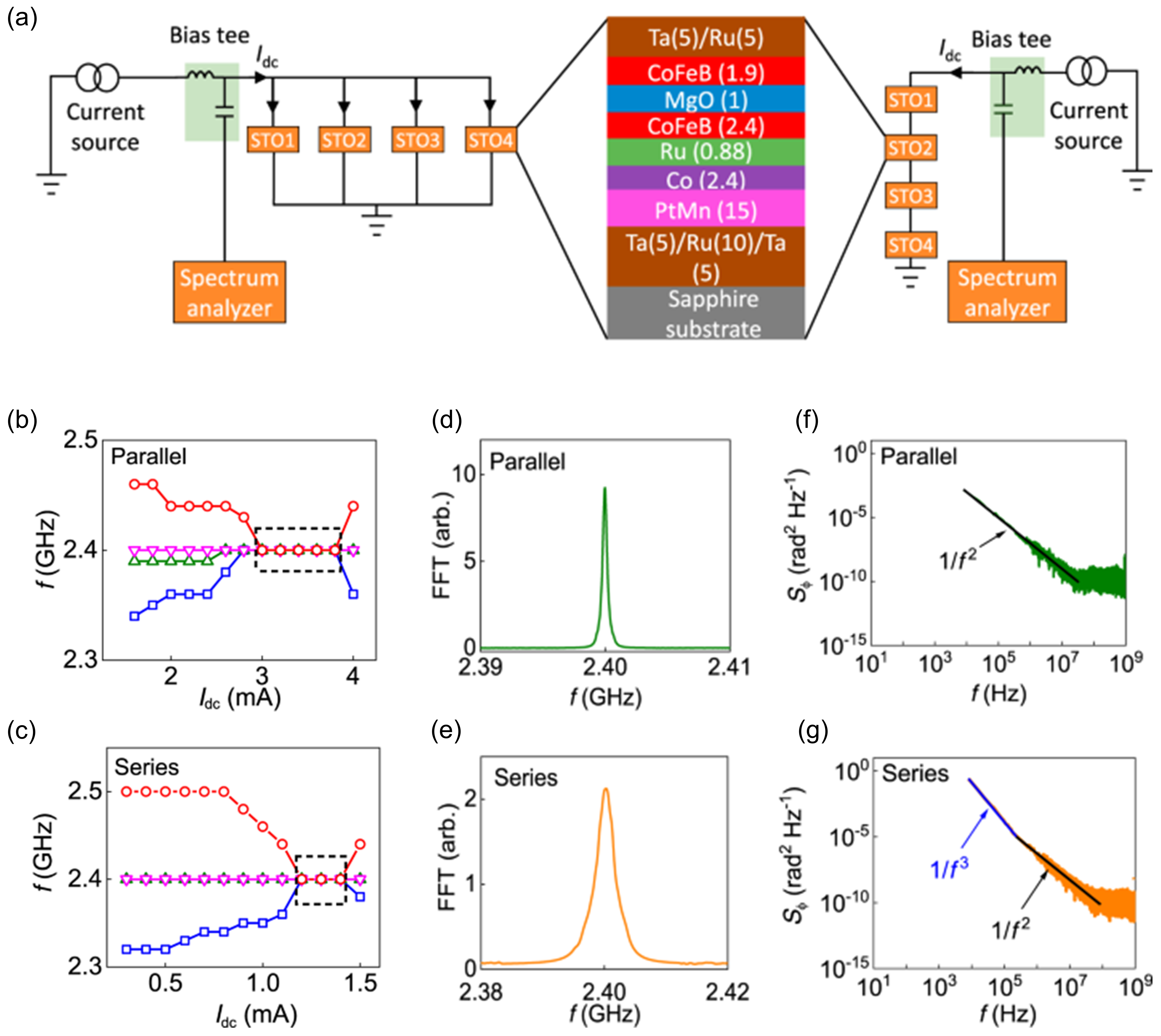}
    \caption{(a) Schematic and measurement set-up for series and parallel connection of STNOs, and the used thin film stacks (numbers in the bracket are the thickness in nm of the individual layers). (b) and (c) Operating frequency versus current of four interacting STNOs connected in parallel and series, respectively. Output spectrum (d) and phase noise (f) for STNOs connected in parallel. Output spectrum (e) and phase noise (g) for STNOs connected in series. Reprinted from~\cite{sharma2021electrically}. }
    \label{fig:Sharma4STNOs}
\end{figure}

Interesting results were obtained recently by Sharma \textit{et al.}~\cite{sharma2021electrically}, demonstrating that depending on the electrical connection scheme, the results may differ significantly. Figure~\ref{fig:Sharma4STNOs}a shows the measurement setup and layout of the investigated STNO connections. % utilized by Sharma et al.~\cite{sharma2021electrically}. 
Since the STNOs are microwave frequency sources, their impedance should be close to 50 $\Omega$ for the best impedance matching with measurement and signal processing blocks. However, depending on the magnetic stack and the STNO design, the optimum resistance for stable and high amplitude magnetization precession may differ from 50 $\Omega$ significantly. In this case, it is convenient to use parallel connections for STNOs with resistance values above 50 $\Omega$ and series connections for STNOs with resistance values well below 50 $\Omega$ \cite{romera2018nt,sharma2021electrically,romera2022binding}. In their experiments, Sharma \textit{et al.}~\cite{sharma2021electrically} found that parallel configuration not only results in a wider locking window (see Fig.~\ref{fig:Sharma4STNOs}b$\&$c) but also results in a larger output power (Fig.~\ref{fig:Sharma4STNOs}d$\&$e). Moreover, it was found that with series connections, the mutual synchronization is weaker because under external injection locking, the serially connected STNOs exhibit 1/$f$ noise below a 200 kHz offset frequency (see Fig.~\ref{fig:Sharma4STNOs}f$\&$g), which is inherent for individual STNOs and indicates the break-up of mutual synchronization.

\section{Spin Hall nano-oscillators}
The simple bilayer structure and in-plane current flow in nano-constriction SHNOs provide large freedom in device design and the layout of oscillator networks. As the nano-constrictions can be fabricated as small as 10 nm, it is possible to also place them very close to each other, which % (close to fabrication limits, i.e. 10 nm) that  
allows for strong dipole-dipole coupling and mode overlap. This makes it possible to synchronize SHNOs in long 1-dimensional chains (both in series and in parallel) and large 2-dimensional arrays. It is also possible to generate propagating spin wave modes in SHNOs~\cite{fulara2019spin}, which will not only allow long-range mutual synchronization but may also provide dynamical and phase-variable coupling between oscillators. SHNOs can also be injection locked to external signals~\cite{demidov2012nm} and it was recently demonstrated that different magneto-dynamical modes injection lock differently as a result of large differences in their mode volume and corresponding energy of the auto-oscillating state~\cite{rajabali2023injection} (see Fig.4 above).
In the following sections, we will discuss the experimental demonstrations of mutual synchronization in SHNOs.

\subsection{Synchronization in chains (1-dimensional arrays)} 

\begin{figure}[t!]
    \centering
    \includegraphics[width=\linewidth]{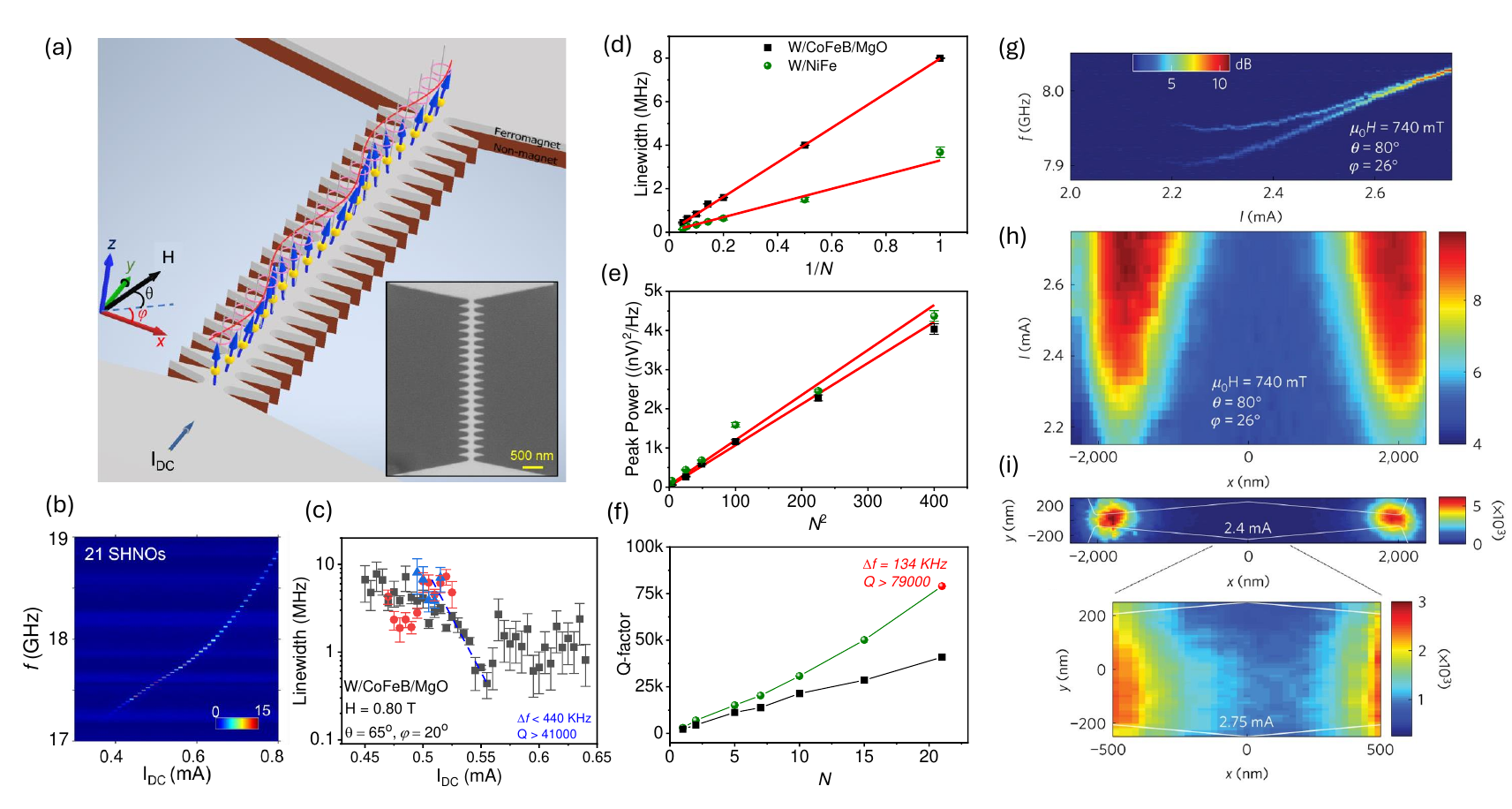}
    \caption{ (a) Schematic of 21 nano-constriction SHNOs connected in series, with an accompanying SEM image shown in the inset. (b) PSD for 21 mutually synchronized oscillators in a chain (for W/CoFeB/MgO thin films, more details can be found in~\cite{kumar2023robust}). (c) The linewidth corresponds to Fig.~8(b). (d-f) spectral linewidth, peak power, and Q-factor ($\Delta$f/f) for varying numbers of oscillators in a chain, respectively. (g-i) PSD and the intensity map using micro-focused Brillouin light spectrometer ($\mu-$BLS) for 2-SHNOs separated by 4 $\mu$m distance. Reprinted with permission from~\cite{kumar2023robust,awad2017natphys}  }
    \label{fig:1Dchains}
\end{figure}

The in-plane current flow in nano-constriction SHNOs allows the fabrication of long chains (1-dimensional arrays) of oscillators by fabricating wider bridges between them. In 2017, Awad \emph{et al.,} reported the first demonstration of mutual synchronization of nine serially connected SHNOs, made from Pt/Py thin films and separated by 300 nm from each other~\cite{awad2017natphys}. The mutual synchronization resulted in lower linewidth (lowest 2 MHz compared to 10 MHz for a single SHNO) and higher output power of up to 54 pW. Recently Kumar \emph{et al.,} extended this robust synchronization to 21 oscillators in a chain with W/CoFeB/MgO and W/NiFe~\cite{kumar2023robust}. Figure~\ref{fig:1Dchains}a shows the schematic and SEM image of the 21 oscillator chain. 
Figure~\ref{fig:1Dchains}b shows the power spectral density of 21 nano-constriction of W(5 nm)/CoFeB(1.4 nm)/MgO(2 nm). The robust mutual synchronization leads to a higher power single excitation. The evolution of the linewidth with current is shown in Fig.~\ref{fig:1Dchains}c. Three different regions of operation were observed: (I) unsynchronized modes with broader linewidth, (II) a decrease in linewidth with the evolution of synchronization between oscillators, and (III) a synchronized state with high linewidth explained as a more chaotic synchronization of oscillators due to higher temperature.  
The robust mutual synchronization leads to a 1/$N$ reduction in linewidth (where $N$ is the number of oscillators in a chain) and an $N^2$ increase in its peak power (shown in Fig.~\ref{fig:1Dchains}d and e). This results in a maximum output power of $>$300 pW and an operational linewidth of $<$134 kHz (quality factor,Q = linewidth/operational frequency = 79000), observed for the W/NiFe system (see Fig.~\ref{fig:1Dchains}f). In the same study, the authors reported single auto-oscillatory modes for up to 50 SHNOs in a chain but with deteriorated linewidth and output power compared to chains with 21 SHNOs. This was explained as due to statistically larger fabrication variation and larger thermal effects in the system, which might be mitigated in the future with material engineering and designing better heat sinks for larger arrays.

The reported mutual synchronization in SHNO chains is mostly driven by dipolar coupling between localized modes in the W/NiFe SHNOs and a combination of dipolar coupling and propagating spin waves in the W/CoFeB SHNOs. In Ref.~\cite{awad2017natphys}, the authors reported synchronization of two nano-constriction SHNOs separated by up to 4 $\mu$m distance (the power spectral density of 2 SHNO separated by 4 $\mu$A is shown in Fig.~\ref{fig:1Dchains}). which was further confirmed with Brillioun light scattering microscopy imaging of the modes (see Fig.~\ref{fig:1Dchains}h and i). This was achieved by making the bridge much more narrow such that at higher currents, a large part of the bridge is also auto-oscillating. These results highlight the long-range mutual synchronization in these devices mediated by direct exchange interaction.

\subsection{Synchronization in two-dimensional arrays}

\begin{figure}
    \centering
    \includegraphics[width=\linewidth]{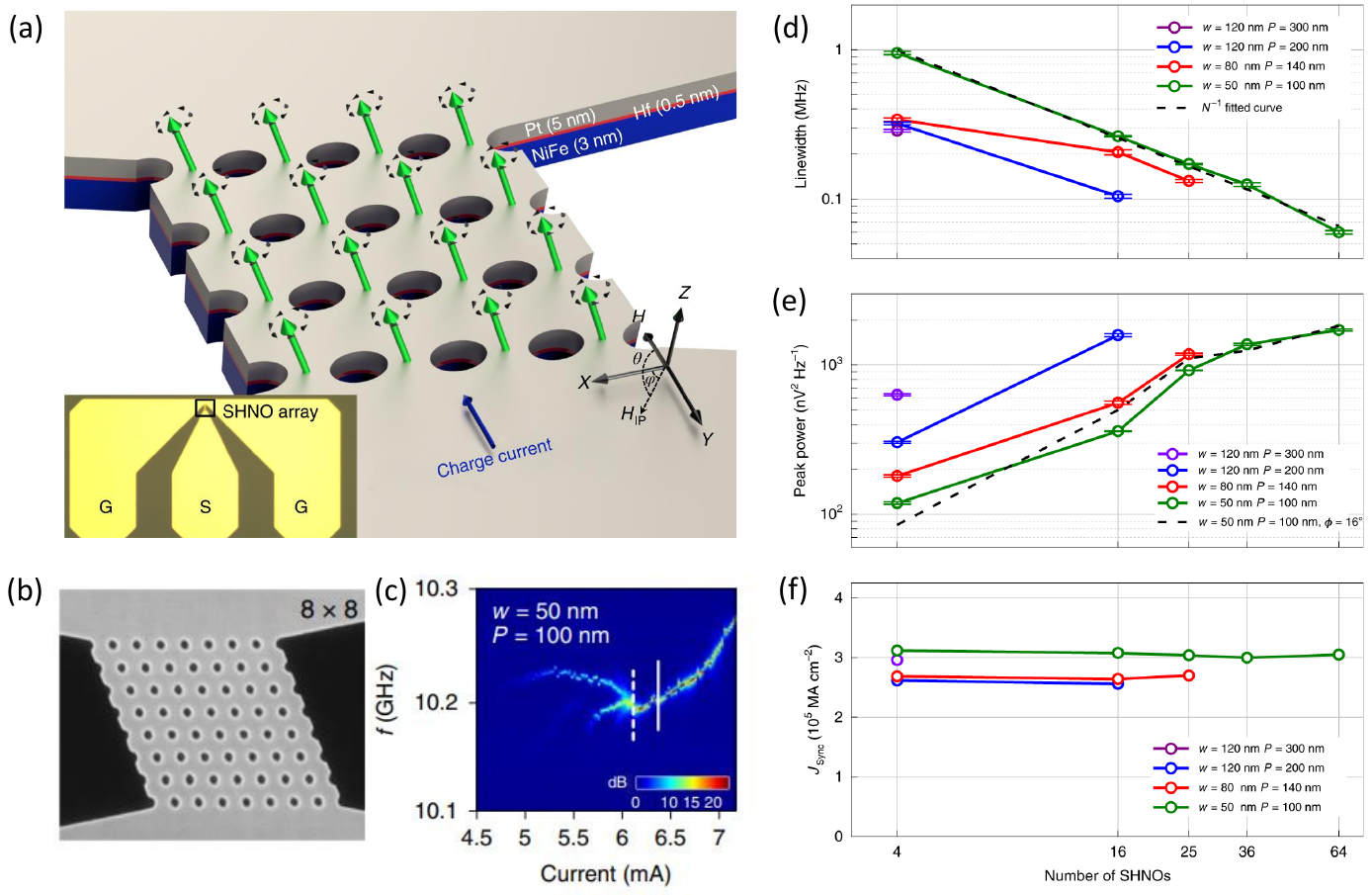}
    \caption{ (a) Schematic of a two-dimensional SHNO array. The inset shows a micrograph of a device with contact pads. (b) The SEM image and (c) PSD of $8\times8$ SHNO array with 50 nm of width and 100 nm centre-to-centre separation. (d-f) Spectral linewidth, peak power, and current density for mutual synchronization for varying sizes of the two-dimensional arrays, respectively. Reprinted with permission from~\cite{zahedinejad2020nt}  }
    \label{fig:2Darrays}
\end{figure}

Nano-constriction SHNOs can also be synchronized in two dimensions (2D). In 2020, Zahedinejad \emph{et al.,}~\cite{zahedinejad2020nt}, fabricated square 2D arrays of up to 10$\times$10 Pt/NiFe nano-constrictions and demonstrated complete synchronization of up to 64 SHNOs (8$\times$8) and partial synchronization of 100 SHNOs. Figure~\ref{fig:2Darrays}a shows the schematic of a 4$\times$4 SHNO array of oscillators fabricated. by simply designing circular holes (of the same order as the oscillator widths) in otherwise extended thin film samples. Figure~\ref{fig:2Darrays}b refers to the scanning electron microscope image of the 8$\times$8 oscillator network. Different nano-constriction widths of $w=$ 50--120 nm and different nano-constriction center-to-center spacing of $p=$ 100--300 nm were investigated. It was found that the number of mutually synchronized SHNOs increased strongly with the inverse of their separation, \emph{i.e.}~in proportion to the dipolar coupling strength. The power spectral density of the auto-oscillating signal from the 8$\times$8 array ($w=$ 50 nm, $p=$ 100 nm) is shown in Fig.~\ref{fig:2Darrays}c. Figure~\ref{fig:2Darrays}d-f summarizes the linewidth, peak power and threshold current density for varying 2D array sizes from 2$\times$2 to 8$\times$8 oscillators. A lowest operational linewidth of 60 kHz and highest peak power of approximately 1800 (nV)$^2$ Hz$^{-1}$ (see Fig.~\ref{fig:2Darrays}d and e) were observed. Though they observed a 1/$N$ reduction in linewidth with the number of SHNOs in an array, the peak power scaled as $N^2$ only for small arrays ($N=$ 4--25). For larger arrays the peak power leveled off, explained as a notable relative phase shift between oscillators farthest apart in the larger arrays. The current density required for mutual synchronization is found to be similar for all arrays (Fig.~\ref{fig:2Darrays}f). 

With recent demonstrations of energy efficient and ultra-narrow SHNOs~\cite{behera2023ultra, behera2022energy}, the number of mutually synchronized SHNOs should be possible to push much higher with the prospect of exceptional spectral parameters and the potential for densely packed populations of SHNOs for emerging computing applications.

\section{External control of mutual synchronization}

For many emerging oscillator applications, direct control of individual oscillators in large oscillator networks will be required. The rich and tunable non-linear dynamics of spintronic oscillators allow individual control of transient mutual synchronization that could be exploited for various applications, including unconventional computing discussed in Sec. 8. Electrical control of mutual synchronization in STNOs was achieved by manipulating the individual bias current in each oscillator that can, in turn, control their operating frequency in oscillators connected in series or in parallel~\cite{lebrun2017mutual,romera2022binding}. The electrical synchronization locks the oscillators when they are within their mutual locking range similar to forced synchronization using injection locking. While this approach has been extended to up to eight STNOs~\cite{tsunegi2018scaling}, it also required individual magnetic fields to reduce the dispersion of their operating frequencies, which is prohibitive if scaled to a much larger number of oscillators.

\begin{figure}
    \centering
    \includegraphics[width=\linewidth]{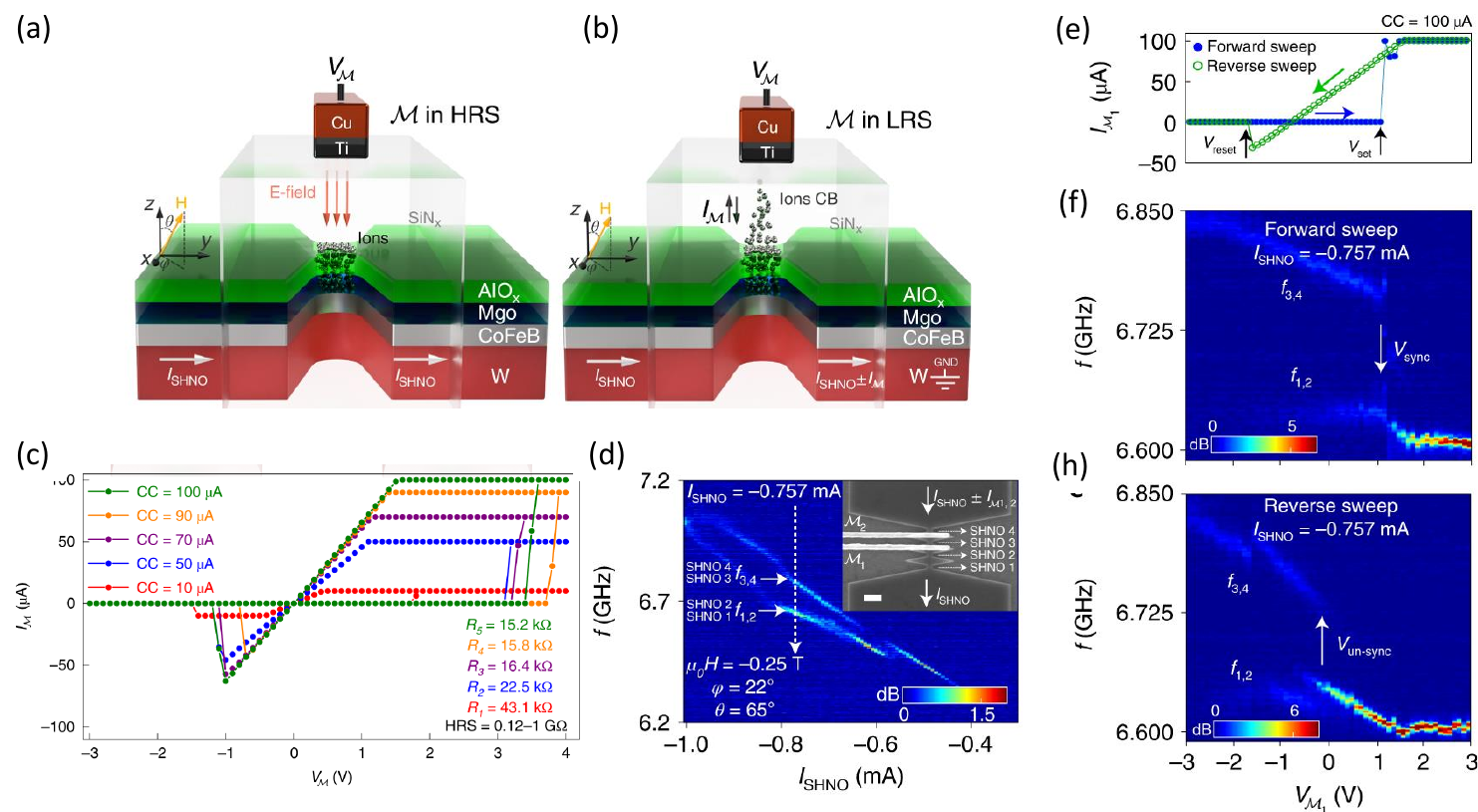}
    \caption{ Schematic representation of memristive gates on top of nano-constriction SHNOs in (a) high resistance state (LRS; E-field control) and (b) low resistance state (HRS; conductive channel). (c) Standard switching curves for memristive elements, where different compliance currents (I$_{CC}$) can provide different LRS values. (d) PSD of 4-SHNOs in a chain. (e) Leakage current characteristic for memristive gate 1 (M$_{1}$) and respective control of mutual synchronization in 4-SHNO chains (f) forward sweep, (h) reverse sweep. Reprinted with permission from~\cite{Zahedinejad2022natmat}  }
    \label{fig:ExControl}
\end{figure}

Using voltage gates fabricated on top of nano-constriction SHNOs, large voltage-controlled frequency tunability of both frequency and threshold current was demonstrated using the voltage controlled magnetic anisotropy (VCMA) effect~\cite{fulara2019spin,kumar2022fabrication,gonzalez2022apl}. Later Choi et al, showed a giant non-volatile frequency tunability exceeding 1 GHz~\cite{choi2022voltage}. This individual control has been extended to chains of four SHNOs to control their mutual synchronization~\cite{Zahedinejad2022natmat} using non-volatile memristive gating. With the application of a gate voltage, metal oxide memristive elements (metal insulator metal materials) undergo an insulating (high resistance state, HRS) to metal (low resistance state, LRS) transition due to the formation of ionic bridges~\cite{mohammad2016nanotech}. Zahedinejad et al.~\cite{Zahedinejad2022natmat} showed that two individually controlled memristor gates can drive the SHNOs chains into different synchronized states (synchronized, partially synchronized, and unsynchronized states). The operation of these devices is shown in Fig.~\ref{fig:ExControl}a and b. In the HRS, the memristors work as a normal voltage gate (generating electric field) and tune the VCMA in a local region. However, when a sufficiently large voltage is applied, the memristor switches to an LRS state and allows the flow of extra current into the SHNO chain, giving global control to SHNOs in a chain. Not only is the memristor non-volatile (even after removing the voltage, the gate remains in its LRS), the resistance of the LRS can be programmed to different values by controlling the size of the conductive bridge by using different compliance current (CC) when forming (see Fig.~\ref{fig:ExControl}c), which also makes the memristors an ideal candidate for memory elements. Figure~\ref{fig:ExControl}d shows the power spectral density for four SHNOs in a chain with two memristive gates M1 and M2 (scanning electron micrograph shown in inset). Figure~\ref{fig:ExControl}e shows the leakage current versus memristive voltage for gate M1. In the forward sweep, when the memristive gate switches from HRS to LRS state ($\approx$ +1 V), an extra (positive) current equivalent to the CC (100 $\mu$A) passes through SHNO 3 and SHNO 4 (shown in inset of Fig.~\ref{fig:ExControl}d), reducing the total current, which lowers the operating frequency of SHNO 3 and SHNO 4, resulting in synchronization with SHNO 1 and SHNO 2 (shown in Fig.~\ref{fig:ExControl}f). Similarly, in the reverse sweep, the memristor switches back to its HRS state (at $\approx$ -1.5 V) and unsyncs the oscillators (see Fig. Fig.~\ref{fig:ExControl}h). In this way these oscillators can be used for on-demand control of mutual synchronization in SHNOs. Recent demonstration of large tunability using memristive nano-gates along~\cite{Khademi2023Large} will provide direct access to each oscillator in 1D chains or 2D arrays. The flexibility of combined voltage and memristive control make this approach scalable to a large oscillator network.  

The recent demonstration of opto-thermal control of SHNOs provides another approach to manipulate its operating frequency and linewidth~\cite{muralidhar2022optothermal}. Using a 532 nm continuous wave laser (spot size of 500 nm and power ranging from 0--12 mW), the authors produced a local temperature rise of about 8 K/mW. With combined current and laser tuning, the authors reported a maximum of 350 MHz tunability in operating frequency (3x larger than current only). With reduced laser spot size (using \emph{e.g.}~nano-plasmonic structures) this could be extended to chains or arrays of oscillators and may provide on-demand temporal or parallel control of oscillators in large networks~\cite{moradi2019spin} without any physical gating.

\section{Benchmarking of spintronic oscillators}
\begin{figure}
    \centering
    \includegraphics[width=0.8\linewidth]{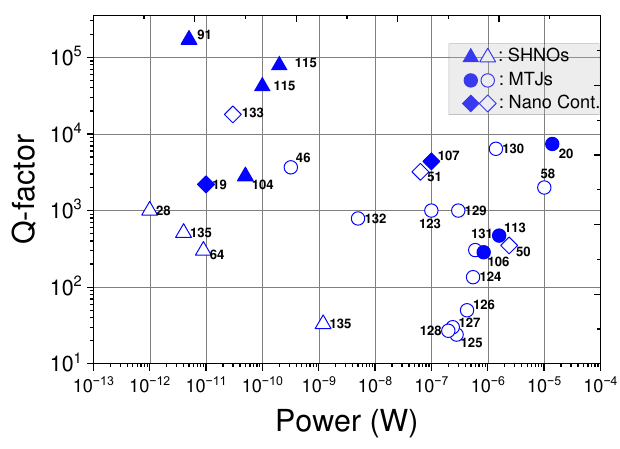}
    \caption{Bench-marking quality factor \emph{vs.}~output power of various spintronic oscillators (hollow symbols) and their synchronized networks (solid symbols). The data comprise the best performance nano-pillar MTJs~\cite{houssameddine2008spin,kubota2013spin,zeng2012high,deac2008bias,zeng2011enhancement,sharma2021electrically,tsunegi2018scaling,costa2017high,seki2014high,lebrun2017mutual}, vortex MTJs~\cite{pribiag2007magnetic,tsunegi2014high,tsunegi2016microwave,dussaux2014large,dussaux2010large}, Nano-contact spin valves~\cite{rippard2004,Maehara2013large,maehara2014high,sani2013mutually,Houshang2015natnano} and SHNOs~\cite{Duan2014,awad2017natphys,fulara2019spin,zahedinejad2020nt,chen2020spin,kumar2023robust}.}
    \label{fig:Benchmarking}
\end{figure}

As discussed in the earlier sections, mutual synchronization of spintronic oscillators leads to larger output power and a more narrow linewidth. For comparison, here we bench-mark different spintronic oscillators with their mutually synchronized counterparts with respect to their quality factor (linewidth/operating frequency) and output power. 
In Fig.~\ref{fig:Benchmarking}, we present a comprehensive compilation of results from previous studies, depicting the $Q$--factor \emph{vs.}~output power for both individual oscillators (hollow symbols) and synchronized oscillators (solid symbols). Throughout this discussion, the following abbreviations are employed: SHNOs for spin Hall nano-oscillators (illustrated by triangles), MTJs for nano-pillar magnetic tunnel junction vortex oscillators (represented by circular symbols), and Nano Contacts for nano-contact spin valve structures (indicated by rhombi).

Observations reveal that vortex oscillators~\cite{pribiag2007magnetic,tsunegi2014high,tsunegi2016microwave,dussaux2010large,dussaux2014large} exhibit the highest output power and are predominantly positioned on the right side of the graph. However, these vortex oscillators typically operate at considerably lower RF frequencies (0.1 GHz to 1.5 GHz), resulting in lower quality factors. Other MTJs and nano-contact devices offer greater frequency tunability but often have lower output power and/or broader operational linewidth, thereby compromising their performance as signal generators.

Synchronized MTJs demonstrate promising results and, as discussed in Section 8 below, are even employed for energy harvesting \cite{sharma2021electrically}. However, they tend to exhibit larger linewidths, translating into lower $Q$--factors. The highest $Q$--factor reported for nano-contact spin valves was approximately 18,000~\cite{rippard2004}, but this corresponded to a suboptimal output power. Recently introduced nano-constriction SHNOs have shown significant potential with their narrow linewidths and high-frequency operation \cite{Demidov2014}. Even individual oscillators exhibit $Q$--factors in the range of 2,000-4,000~\cite{Duan2014,zahedinejad2018cmos,fulara2019spin}, albeit with extremely low output power.

The 2D arrays of 64 SHNOs demonstrate an impressive $Q$--factor of 179,000, albeit in a white noise regime (measured at short time scales). Using synchronization in extended chains, the enhanced output power was observed, achieving up to 200 pW while maintaining a high $Q$--factor exceeding 79,000 (for W/NiFe thin films). The NiFe/Pt system exhibits an even higher output power of 300 pW with a $Q$--factor exceeding 32,000. With the recent demonstration of 10 nm SHNOs \cite{behera2023ultra}, it is predicted that synchronization of much larger numbers of SHNOs will be possible, thanks to the much stronger dipolar coupling at short separations. Therefore, it is likely that even higher $Q$--factors will be achievable.

\subsection{Phase Noise}

\begin{figure}[b!]
    \centering
    \includegraphics[width=\linewidth]{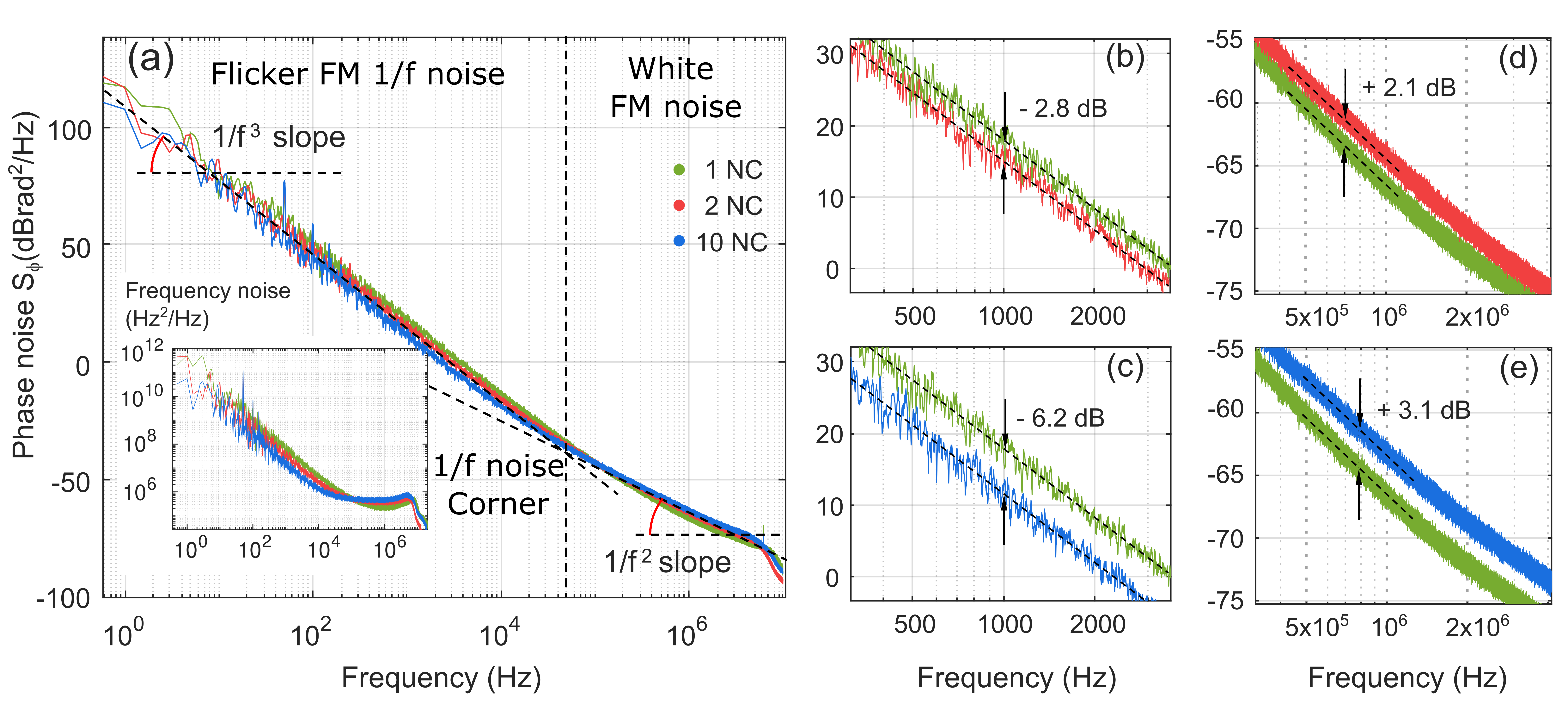}
    \caption{(a) Phase noise plots for single and mutually synchronized NC SHNO chains. (b) Zoomed area of phase noise at low offset frequencies for single and 2 NC SHNOs. (c) Zoomed area of phase noise at low offset frequencies for single and 10 NC SHNOs. (d) Zoomed area of phase noise at high offset frequencies for single and 2 NC SHNOs. (e) Zoomed area of phase noise at high offset frequencies for single and 10 NC SHNOs. Reprinted with permission from ~\cite{litvinenko2023phase} }
    \label{fig:LitvinenkoNCSHNOsPhaseNoise}
\end{figure}

Phase noise plot is a complex measure that has been actively employed for the analysis of oscillator properties and allows to identify physical nature of different contributions to the signal instability. The phase noise plot represents the level of random phase fluctuations at different offset frequencies from the carrier. Generally, the phase noise of free-running oscillators increases to a lower offset frequency at different slopes. The slope of the phase noise curve indicates the physical nature of the noise at the corresponding frequencies. The main classified types of phase noise are white phase or thermal noise with $1/f^0$ slope in the phase noise plot, flicker phase noise with $1/f^1$ slope, white FM or random walk PM with $1/f^2$ slope, flicker FM 1/f noise with $1/f^3$ slope and random walk FM noise with $1/f^4$ slope.

\begin{figure}[t!]
    \centering
    \includegraphics[width=\linewidth]{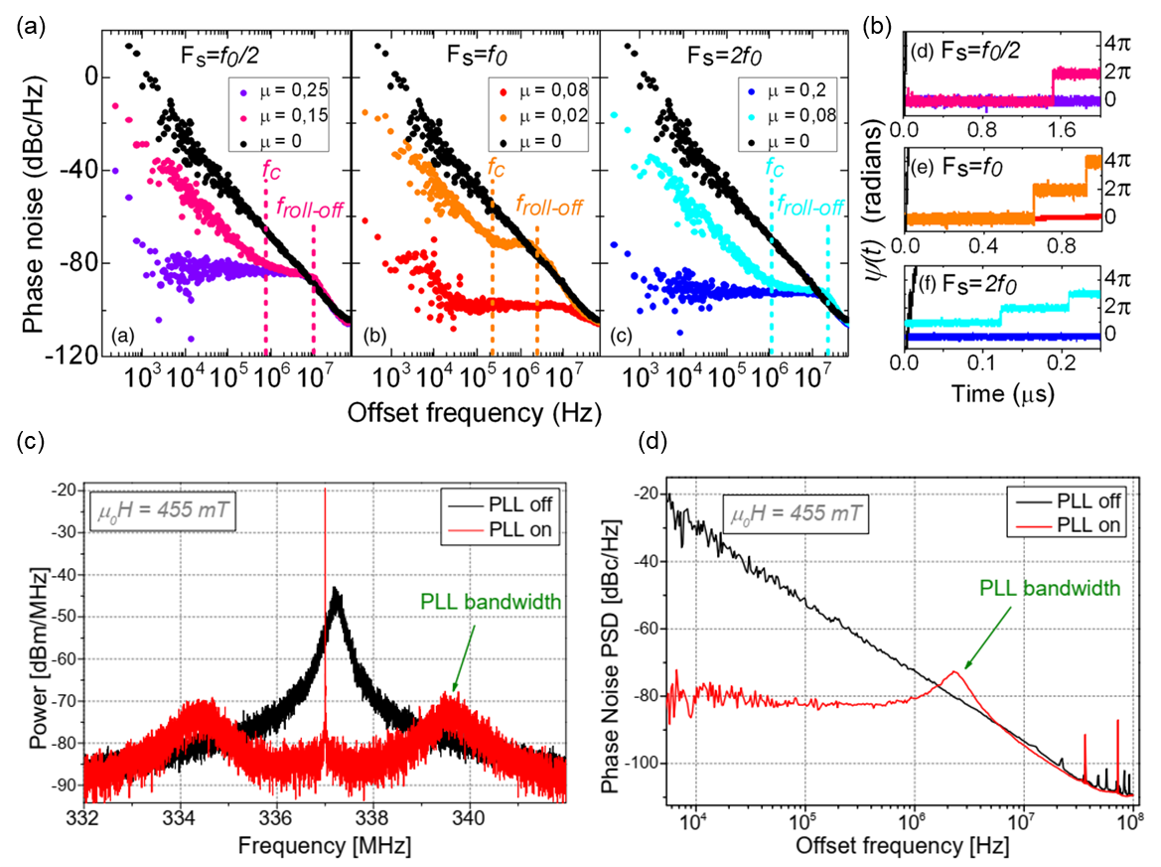}
    \caption{(a) Phase noise plots for $f_{0}/2$, $f_{0}$, and $2f_{0}$ fractional synchronization schemes. (b) The corresponding time traces of the instantaneous phase show occasional phase slips. (c) Power spectra for a free-running STNO (in black) and a PLL-stabilized STNO (in red). (d) The corresponding phase noise plots for a free-running STNO (in black) and a PLL-stabilized STNO (in red). Reprinted with permission from ~\cite{Lebrun2015, wittrock2021stabilization} }
    \label{fig:PhaseNoise}
\end{figure}

A basic theoretical description of the thermal fluctuations can be done using Eq.~\ref{eq:oscillator} and selecting $\xi(t)$ to be white Gaussian noise \cite{kim2008prl, slavin2009nonlinear}. The power spectrum in this approach has a Lorentzian shape with a full linewidth at half maximum of
\begin{equation}
    2\Delta\omega=(1+\nu^2)\Gamma_+\frac{k_B T}{E_0},
\end{equation}
where $k_B$ is the Boltzmann constant, $T$ is the temperature, and $E_0$ defines the energy of the auto-oscillation. The linewidth is quite large since spintronic nano-oscillators are highly non-linear and the energy of oscillation is quite low. The generation linewidth can be reduced by tuning the external field, the anisotropy~\cite{jiang2020reduced}, and the current to a point of minimum or zero nonlinear frequency shift. 

The analysis of phase noise unveils the underlying physics that leads to weaker frequency stability and larger integral linewidth. In~\cite{litvinenko2023phase}, researchers demonstrated a sufficient improvement of the phase noise in the 1/$f$ region (Flicker FM) of phase noise for a chain of 10 mutually synchronized SHNOs. However, in the White FM region of phase noise, a significant degradation was observed. This effect can be associated with several factors, such as process variation of the nano-constrictions, temperature gradients within the chain making the SHNOs nonidentical and increasing the overall temperature, and, finally, phase delays in the coupling between nano-constrictions, which may lead to decoherence in the chain and elevated noise levels.

The use of external synchronization ~\cite{Lebrun2015} improves the phase noise significantly, as shown in Fig.~\ref{fig:PhaseNoise}a,b. The phase noise plot exhibits a plateau starting from a roll-off frequency and down to a corner frequency, where it again rises. The roll-off frequency is defined by the amplitude relaxation frequency under external synchronization. The growing phase noise below the corner frequency and closer to zero offset frequencies is associated with occasional phase slips. Once complete synchronization and phase locking are achieved, the phase slips disappear and the phase noise flattens till the lowest offset frequencies.

Similarly, the use of external stabilization systems, such as phase locked loops (PLL)~\cite{wittrock2021stabilization}, has a similar effect on phase noise, effectively flattening it below the PLL bandwidth frequency offset (see Fig.~\ref{fig:PhaseNoise}c,d). Both methods of stabilization can also be exploited for mutually synchronized nano-oscillators.
\cite{tsunegi2018scaling,litvinenko2023phase}

\section{Applications}
The improvement in operational parameters (output power, linewidth, and frequency tunability) and connectivity/interaction between oscillators in a network of synchronized oscillators promote these oscillators for a wide range of applications~\cite{stamps2014jphysd,dieny2020natelectron}. In the following subsections, some of the promising applications are discussed in detail.

\subsection{Signal generation and analysis}

\subsubsection{Wireless communication}

\begin{figure}[b!]
    \centering
    \includegraphics[width=\linewidth]{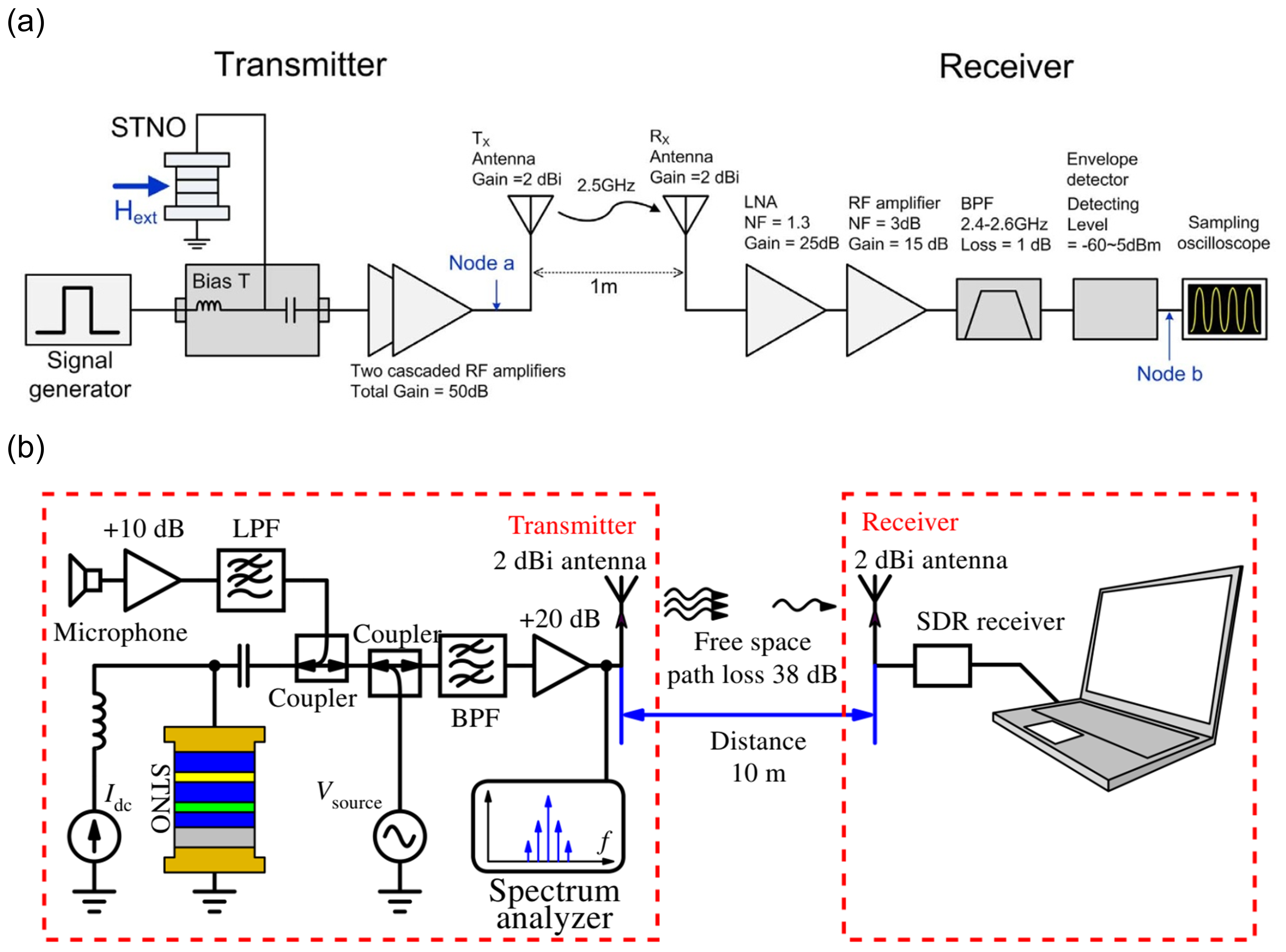}
    \caption{(a) Schematic of a data transmission system based on ASK modulation of an in-plane magnetized STNO. (b)Schematic of a data transmission system based on PM and PSK modulation of an out-of-plane magnetized vortex-state STNO. Reprinted with permission from ~\cite{litvinenko2021analog,choi2014spin} }
    \label{fig:DataTransmission}
\end{figure}

Since the first demonstration of spintronic oscillators, there were many attempts to design STNO- and SHNO-based wireless communication schemes \cite{manfrini2009apl,manfrini2011jap,pufall2005,muduli2010,pogoryelov2011a,pogoryelov2011b,Choi2014scirep,litvinenko2021analog,Sharma2014apl}, with the hope of harnessing their high operational frequency, wideband frequency tunability, and fast frequency tuning, which are of utmost interest for wireless sensor network (WSN) and Internet of Things (IoT) communication systems. Basic modulation schemes were demonstrated, such as frequency and amplitude modulation (FM and AM) \cite{manfrini2009apl,manfrini2011jap,pufall2005,muduli2010,pogoryelov2011a,pogoryelov2011b,pogoryelov2012combined,sharma2021electrically,tsunegi2018scaling} achieving up to 10 MHz modulation rates for relatively low-frequency vortex-state STNOs and up to 500 MHz for high-frequency (24.5 GHz) uniform-state GMR-based STNOs, which promises data rates of up to 1 Gb/s. Moreover, the demonstration of high-rate FM was performed with a pair of synchronized high-frequency nanocontact STNOs. The use of a synchronized STNO pair improved the signal output power by 3 dB and, most importantly, the study demonstrated that the STNO pair keeps synchronized even under a large modulation current and a fast modulation rate of 500 MHz. However, due to the large phase noise and linewidth combined with the low output power of spintronic oscillators, their exploitation in transmitter circuits limits them to the simplest binary modulation schemes, which consumes a large bandwidth and makes spintronic-based FM and AM communication solutions incompatible with conventional communication protocols.

Later, phase modulation (PM) and phase shift keying (PSK) methods were adopted for relatively low-frequency vortex-based STNOs \cite{litvinenko2021analog}. The key ingredient of the PSK and PM methods is the use of an external frequency locking mechanism of the nonlinear STNOs in order to improve the phase noise and reduce the linewidth to 1 Hz. The phase modulation is achieved by changing the intrinsic (free-running) frequency via injecting an additional low-frequency current through the STNO. Since the demonstration was performed with low-frequency vortex-based STNOs, the maximum data rate was 5 Mb/s. However, the use of high-frequency spintronic oscillators promises a sufficient increase of up to 500 Mb/s -- 2 Gb/s in data rates.

%\subsubsection{LoRa}

\subsubsection{Spectrum analysis}
Another application where mutual synchronization could improve system parameters is ultra-fast wide-band spectrum analysis (UFSA). Initially, this concept was implemented with STNOs in a single vortex and in a uniform state, demonstrating how to exploit their very fast frequency tunability~\cite{litvinenko2020ultrafast,litvinenko2022ultrafast}.

\begin{figure}
    \centering
    \includegraphics[width=\linewidth]{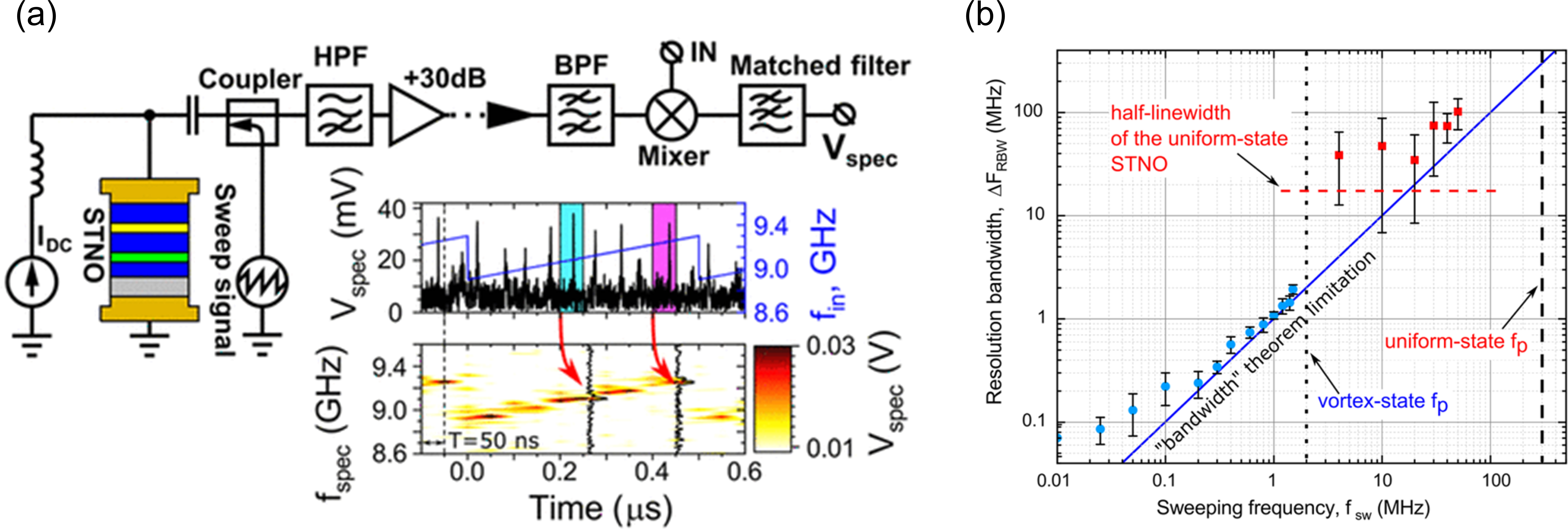}
    \caption{(a) Schematic of an ultra-fast spectrum analyzer (UFSA) based on a uniform-state STNO. The top inset shows time traces of the output signal. The bottom inset shows a spectrogram obtained with a 50 ns sweep time for an input signal of 9.1 GHz central frequency with FM up-chirp modulation with 500 ns modulation period and 500 MHz modulation amplitude. (b) Resolution bandwidth for a uniform-state STNO-based UFSA and a vortex-state STNO-based UFSA as a function of sweeping frequencies. Reprinted with permission from ~\cite{litvinenko2022ultrafast} }
    \label{fig:UFSA}
\end{figure}

The idea of UFSA is based on mixing the input analyzed signal with a signal of a rapidly swept-tuned nano-oscillator and the subsequent extraction of the spectra with a matched filter. When the tuning speed of the nano-oscillator is in the range of $\mu$s to ns and the input frequency within this period can be considered constant, frequency mixing serves as a Z-transform operation for the input signal. The matched filter performs an inverse Z-transform and gives a pulsed signal where each pulse with its amplitude and temporal position corresponds to the spectrum of the input signal.

It was demonstrated for vortex-state and uniform-state STNOs that at fast sweeping frequencies, the phase noise of the nano-oscillators has no contribution to the resolution bandwidth (RBW) of the UFSA since at $\mu$s and ns time scale, the signal from the STNOs can be considered monochromatic (see Fig.~\ref{fig:UFSA}e). However, at low offset frequencies, the RBW does decrease due to the phase noise of the STNOs. In \cite{litvinenko2022ultrafast}, researchers projected that the use of STNOs with narrower generation line width would bring the frequency resolution of UFSA even closer to the limit imposed by the bandwidth theorem even for low scanning frequencies. Hence, the use of large arrays of mutually synchronized SHNO could potentially allow for sufficient expansion of the range of sweeping frequencies at which the resolution bandwidth will be mostly defined by the bandwidth theorem.

\subsection{Unconventional Computing}

Coupled oscillator systems exhibit significant potential for unconventional computing, as highlighted in previous research~\cite{csaba2020coupled}. The nonlinear dynamics of spintronic oscillators, along with mutual synchronization to generate binding events, render these devices suitable for diverse computing schemes. In this context, we review the experimental demonstrations of neuromorphic and reservoir computing implementations leveraging the synchronization properties of STNOs and SHNOs. Additionally, we explore the utilization of these devices for implementing Ising machines.

\subsubsection{Neuromorphic and Reservoir Computing} 

\begin{figure}[b!]
    \centering
    \includegraphics[width=\linewidth]{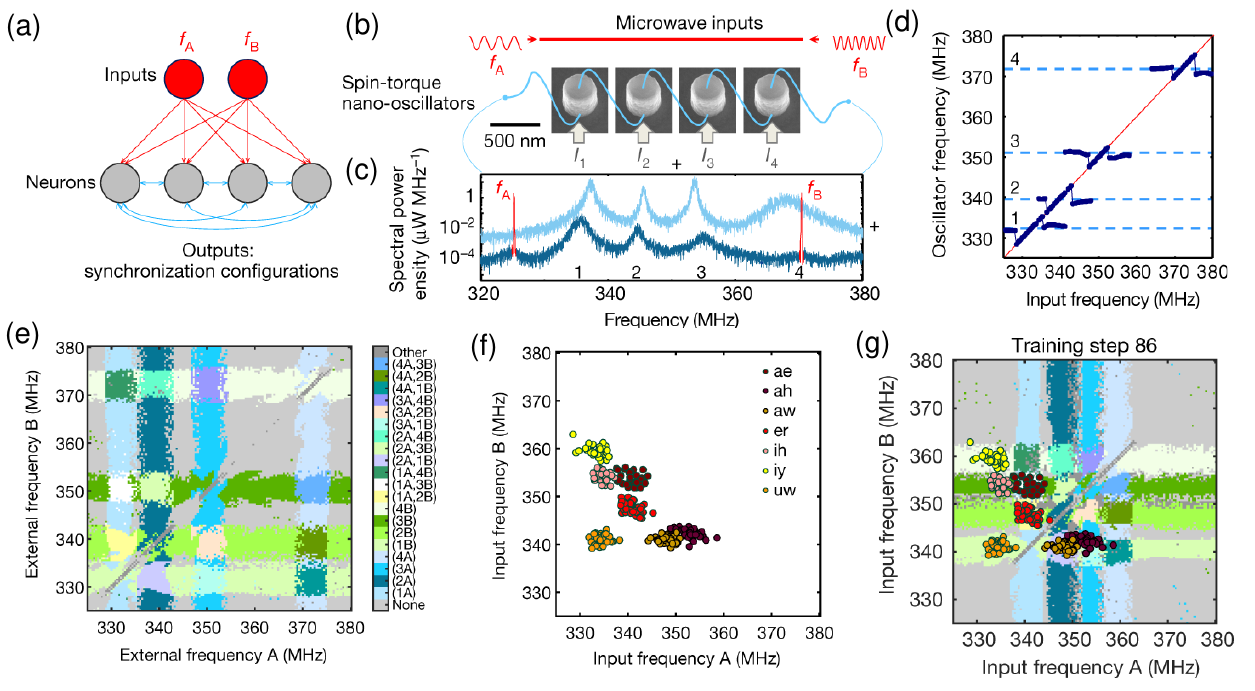}
    \caption{ (a) Schematic of a neural network with four neurons connected with each other. (b) Experimental configuration with four STNOs connected in series and electrically coupled by their output microwave power. The French vowels are encoded into two microwave signals (f$_A$ and f$_B$) applied to the oscillator networks using microwave field lines. (c) The measured microwave signal from the oscillator networks in the absence (light blue) and presence (dark blue) of encoded signals. (d) Injection locking pattern of four intrinsic microwave signals (from oscillator network) with input frequency in the field line. (e) Synchronization map of external frequencies (f$_A$ and f$_B$). (f) Inputs applied corresponding to a different spoken French vowel, the data points represent different speakers. Reprinted with permission from ~\cite{romera2018nt} }
    \label{fig:ReservoirComp}
\end{figure}

The fact that neurons in brain behave as non-linear oscillators has motivated researchers to look into coupled oscillator dynamics for unconventional computing applications~\cite{csaba2020coupled}. Because of their nanoscopic size, spintronic oscillators are one of the primary choices for ultra-compact neuromorphic hardware. Furthermore, just as a neuron reacts to the spiking of neighboring neurons, individual oscillators respond to external stimuli. Torrejon \textit{et al.} demonstrated the first spintronic oscillator-based neuromorphic computing scheme (using vortex STNOs) with accuracy similar to state-of-the-art neural networks~\cite{torrejon2017nt}. Using the amplitude dynamics of a single STNO with time-multiplexing (to emulate a full neural network), they achieved spoken digit recognition of 99.6\% (with cochlear filtering). Later, Romera et al.~\cite{romera2018nt} demonstrated reservoir computing using four STNOs (vortex state MTJs with 375 nm diameter and FeB as the free layer) connected in series. Figure~\ref{fig:ReservoirComp}a shows the schematic of the emulated neural network. Since the number of oscillators was limited only 2 external inputs in the form of injection locking frequencies were utilized in the demonstration. Figure~\ref{fig:ReservoirComp}b depicts the experimental setup with four STNOs connected in series and electrically coupled through their emitted microwave current. Here, the microwave frequency of each oscillator is controlled by individual currents in each STNO (I$_{1}$, I$_{2}$, I$_{3}$ and I$_{4}$). Figure~\ref{fig:ReservoirComp}c, the four broad peaks in light blue show the intrinsic free-running generation peaks of these oscillators. The two narrow red peaks correspond to the microwave signals associated with spoken vowels. Torrejon \textit{et al.} used a subset of the Hillenbrand database comprising seven vowels pronounced by 37 different female speakers, where each vowel is characterized by 12 different frequencies. Formant frequencies from the range between 500~Hz and 3500~Hz were mapped onto input frequencies (f$_{A}$, f$_{B}$) in the range between 325~MHz and 380~MHz which corresponds with a possible tuning range of all the oscillators using linear combination of the 12 formant frequencies that fit the grid-like geometry. Figure~\ref{fig:ReservoirComp}d shows the behaviour of the instantaneous frequency of all four individual oscillators as the function of frequency of a single injection locking signal. The range of frequencies where the frequencies of oscillators follow the external input signal correspond to the injection locking regions. Similarly, with two external signals an grid-like synchronization 2-D map can be plotted. Figure~\ref{fig:ReservoirComp}e shows the experimental synchronization map for external frequency A and frequency B with STNOs. Each color represents their locking in a different synchronization state e.g. (4A,3B) representing the injection locking of the 4th STNO with signal A, and 3rd STNO with the input signals B, respectively. The associated vowel recognition map is shown in Fig.~\ref{fig:ReservoirComp}f.
The principle of recognition is the following - the individual currents of all 4 oscillators are adjusted so that maximum a group of two arbitrary oscillators demonstrate an independent injection locking at the same time, each oscillator to a corresponding input frequency A or B (see Fig.~\ref{fig:ReservoirComp}g). This allows to exploit oscillators injection locking as a detection map for two external frequencies. The fact of injection locking can be easily and rapidly determined with a simple amplitude detector circuit.

 A similar approach was demonstrated by Zahedinezad et al.~\cite{zahedinejad2020nt} using a 2D array of synchronized SHNOs. The authors used a $4\times4$ oscillator array to improve the linewidth of each $4\times1$ sub-array with mutual synchronization and improve the recognition rate. Then each $4\times1$ sub-array was used as an individual detecting oscillator to demonstrate vowel recognition by injection locking the intrinsic frequency (of each $4\times1$ sub-array) to two input frequencies associated with the vowels.

\begin{figure}[b!]
    \centering
    \includegraphics[width=\linewidth]{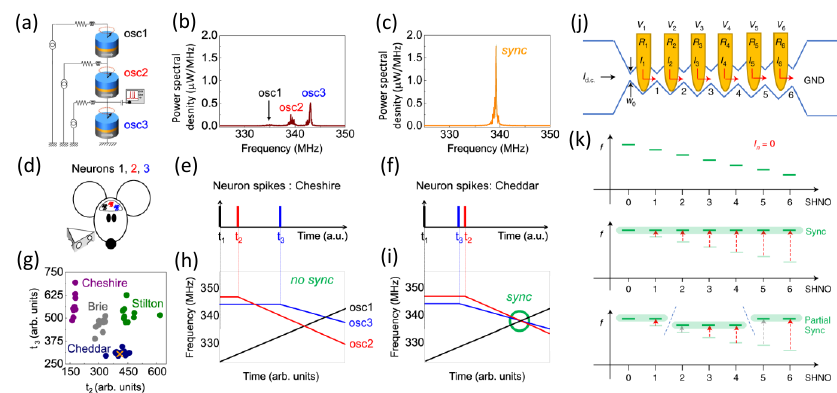}
    \caption{(a) Schematic of three STNOs connected in series and coupled with their electrical microwave currents. (b) The measured microwave signal from the oscillator network in an unsynchronized and (c) synchronized state. (d) Schematic of a fictitious mouse with three neurons. Different cheese presented to the mouse generates different responses in the mouse's brain. (e) and (f) Example of neuron spikes in the mouse brain for Cheshire and Cheddar cheese, respectively. (g) Applied inputs, represented as the time of spike for neurons 2 and 3 (the time of spike from neuron 1 is considered zero to set at origin of sequence). (h) and (i) Current ramps corresponding to the applied input sequences shown in (e) and (f), respectively. (j) Schematic representation of an SHNO chain with 7 nano-constrictions (0-6) of incremental width, where each connecting bridge (between nano-constrictions) consists of a memristor gated to add extra current in the forward SHNOs. (k) (top to bottom) Initial microwave signal from individual SHNOs, microwave signal in the synchronized state (by adding additional currents from the memristors), a partially synchronized state using different ON and OFF states in the memristors. Reprinted with permission from ~\cite{romera2022binding,Zahedinejad2022natmat} }
    \label{fig:NeuromorphicComp}
\end{figure}

Romera \textit{et al.}~\cite{romera2022binding} exploited the ability of mutual synchronization in STNOs for temporal pattern recognition by binding consecutive events in time. In their recent demonstration, the authors utilized the non-linear dynamic properties of oscillators in a network for precise control of mutual synchronization by tuning the oscillator frequencies over a wide range using input charge current. Figure~\ref{fig:NeuromorphicComp}a shows the schematic of the oscillator network of three STNOs connected in series. Here the DC biasing voltages controlled in a such way to provide independent control of injected current levels for all three oscillators. %Here the current fed to the first oscillator flows through all STNOs, the current given to the second oscillator adds to oscillator 2, and oscillator 3 has its own current that does not add to any other STNO. 
Figure~\ref{fig:NeuromorphicComp}b shows the independent microwave emission from each oscillator. The series interconnection allows the AC microwave signals from each STNO to act on other STNOs. The operating frequency of individual STNOs can be manipulated using individual DC currents and brought within their locking range, which results in complete synchronization (PSD shown in Fig.~\ref{fig:NeuromorphicComp}c). To test their system, the researchers formed a dataset for four different types of cheese (Cheshire, Brie, Cheddar, and Stilton) presented to a fictitious mouse with three neurons (schematically shown in Fig.~\ref{fig:NeuromorphicComp}d). The sequence of spikes for Cheshire and Cheddar cheese are shown in Fig.~\ref{fig:NeuromorphicComp}e $\&$ f. The first spike is considered as origin and hence, its time is taken as zero. The time sequence for the other two spikes corresponding to different cheeses are shown in Fig.~\ref{fig:NeuromorphicComp}g (different colors represent different cheese and different points show samples of cheese per category). The objective of the task is to recognize the cheese by providing a particular time sequence. The time sequence for each category is then used to trigger a decrease in individual DC control currents with a specific for each STNO temporal slope, which results in different temporal synchronization pattern between STNOs. % to individual currents in the STNOs, which results in their distinguishable synchronization pattern. 
The time sequence of Cheshire cheese is shown in Fig.~\ref{fig:NeuromorphicComp}h, where first we can observe a synchronization between STNO 1 and STNO 2 with a following synchronization between STNO1 and STNO3. %the ramps do not intersect with each other and hence can be recognized if the STNOs do not synchronize with each other. 
On the other hand, the time sequence of Cheddar cheese (shown in Fig.~\ref{fig:NeuromorphicComp}i) results in the intersection of all three individual frequencies (defined by individual current values for each STNO) at the same time and results in their complete mutual synchronization. %sintersects at a specific time (set of current values for STNOs) and results in complete mutual synchronization. 
The demonstration achieved an overall 94$\%$ recognition rate compared to 93.3$\%$ for a perceptron trained on the same database. 
Recently, using low-power memristive control of mutual synchronization of SHNOs in a chain, Zahedinejad \textit{et al.}~\cite{Zahedinejad2022natmat} also demonstrated the potential of SHNO networks for programmable ultra-fast pattern recognition. The schematic for the proposed configuration is shown in Fig.~\ref{fig:NeuromorphicComp}j, where seven nano-constrictions (0-6) with incremental widths are connected in series. The incremental width not only makes their threshold current higher for each upstream neighbouring SHNO but also decreases the microwave frequency (shown in Fig.~\ref{fig:NeuromorphicComp}k). Hence, without any additional input, the chain of SHNOs results in an unsynchronized microwave signal from each nano-constriction. Using a memristor (in their LRS) on the bridge connecting the nano-constrictions, one can manipulate the state of synchronization state in the chain by adding or subtracting the additional current. In this way, the initially unsynchronized SHNOs (due to differences in their widths) can be synchronized with additional current provided by memristive gates. As per their demonstration, a large chain of nano-constriction SHNOs with varying widths of nano-constrictions can be driven into an unsynchronized, partially synchronized, and completely synchronized state by varying the memristor states. These synchronization states can be distinguished by their significantly different output power.

\subsubsection{Ising Machines} 
Ising machines (IMs) are analog and digital physical computing systems designed to rapidly find solutions to combinatorial optimization (CO) problems. IMs are programmed via the coupling strengths between its binarized spins. The advantage of IMs is that they intrinsically tend to minimize their energy by relaxing over time to their lowest-energy state, which is the solution to the CO problem. Ising machines are characterized with a square root pre-factor advantage over existing algorithms for CO optimization, i.e time-to-solution scales as $\exp(c \sqrt{N})$. IMs have been implemented on different platforms, where interacting nonlinear high-frequency oscillators are particularly promising candidates \cite{albertsson2021ultrafast,houshang2022prappl,mcgoldrick2022ising,litvinenko2023spinwave} since their relaxation time is in nanoseconds and, therefore, their time-to-solution is on the order of microseconds.

\begin{figure}
    \centering
    \includegraphics[width=1\linewidth]{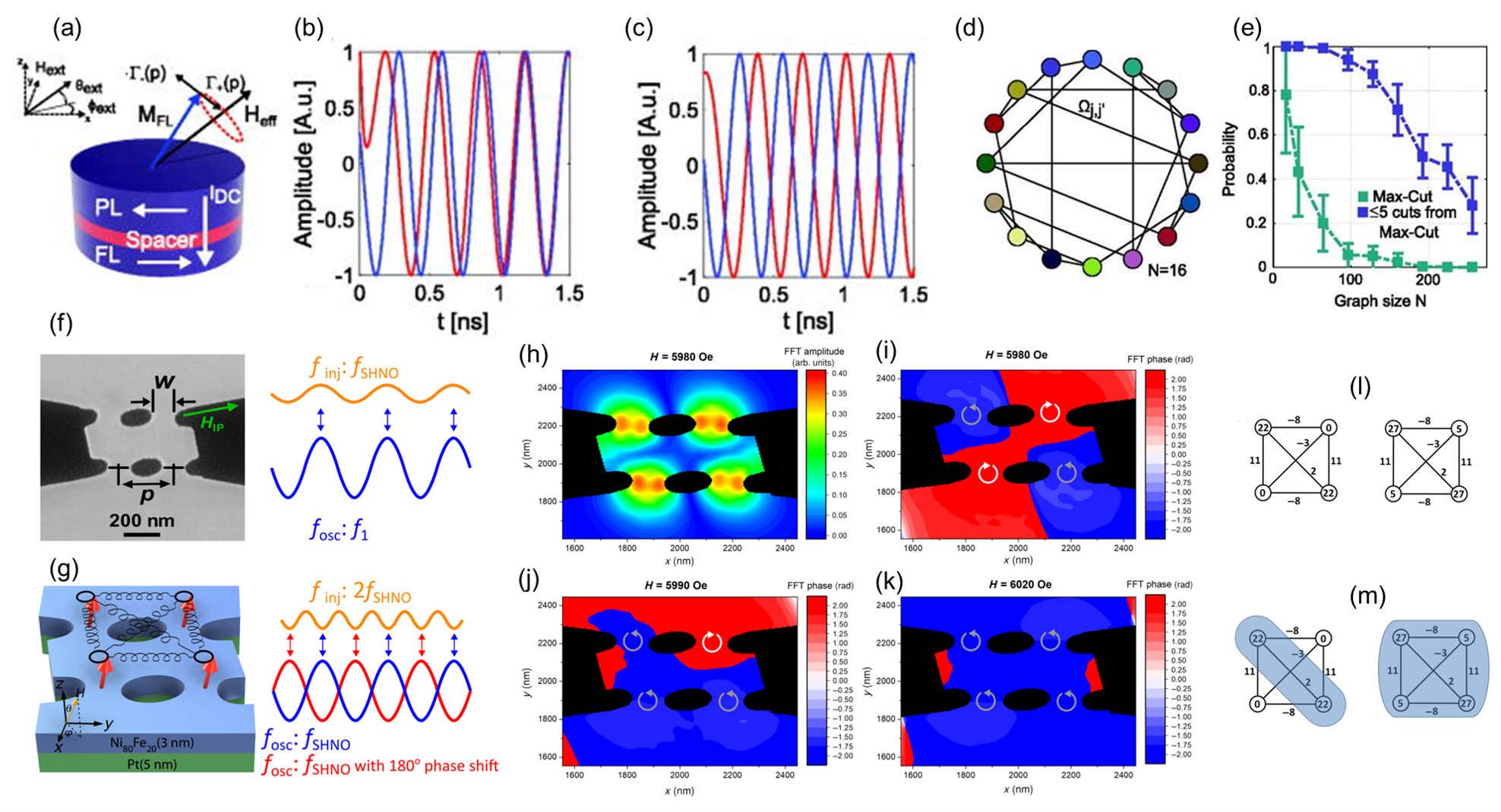}
    \caption{(a) The STNO and its configuration.  (b,c) Two stable phase-binarized states of an STNO pair when both oscillators are (b) in-phase and (c) out-of-phase. (d) MAX-CUT problem with arbitrary interconnection. (e) Probability for an optimum ground state solution and sub-optimum solution with 5 smaller cuts number from an optimum. (f,g) A 2$\times$2 SHNO array with a demonstration of single phase at $f$ and phase binarization at 2$f$ synchronization. (h) Amplitude of the magnetization and phase (i) when 4 oscillators are synchronized diagonally. (j) Simulated phase maps when 3 oscillators are in-phase and one oscillator is out-of-phase. (k) Phase map when all oscillators are in-phase. (l,m) Corresponding MAX-CUT configurations. Reprinted from~\cite{albertsson2021ultrafast,houshang2022prappl}}
    \label{fig:IMd}
\end{figure}

The idea of using oscillators for Ising machines is based on the phenomena of external fractional and mutual synchronization~\cite{gonzalez2024spintronic,kumar2024spin}. When an oscillator is synchronized to an external source at twice its intrinsic frequency it gets stabilized at either 0 or 180 degrees (see Fig.~\ref{fig:IMd}b,c,f,g) relative to the phase of an external source, which makes it a perfect analog of an Ising spin. As was demonstrated in \cite{albertsson2021ultrafast}, spintronic oscillators can be effectively exploited to build an Ising machine and solve simple MAX-CUT problems with up to 250 spins. In order to operate effectively with a large number of oscillators, oscillator-based IMs should have good mutual and external synchronization properties, \emph{i.e.}~an oscillator can be effectively phase binarized under external synchronization and can change their state when weakly and negatively coupled to another oscillator with the same state or a group of oscillators with the same state. Coupling between STNOs can be implemented electrically with cables \cite{lebrun2017mutual,tsunegi2018scaling} or with magnetic dipolar fields \cite{Locatelli2015}.

The use of SHNOs in IM design also allows for synchronization between oscillators via propagating spin waves. Moreover, SHNOs have much higher oscillation frequencies that reduce the time-to-solution to below microseconds. Figure~\ref{fig:IMd}f-m shows the construction of an SHNO-based IM and successful experimental demonstration of phase binarized SHNOs solving a 4-spin MAX-CUT combinatorial problem~\cite{houshang2022prappl}.

\subsection{Energy Harvesting and Signal Detection} 
\begin{figure}[h!]
    \centering
    \includegraphics[width=0.6\linewidth]{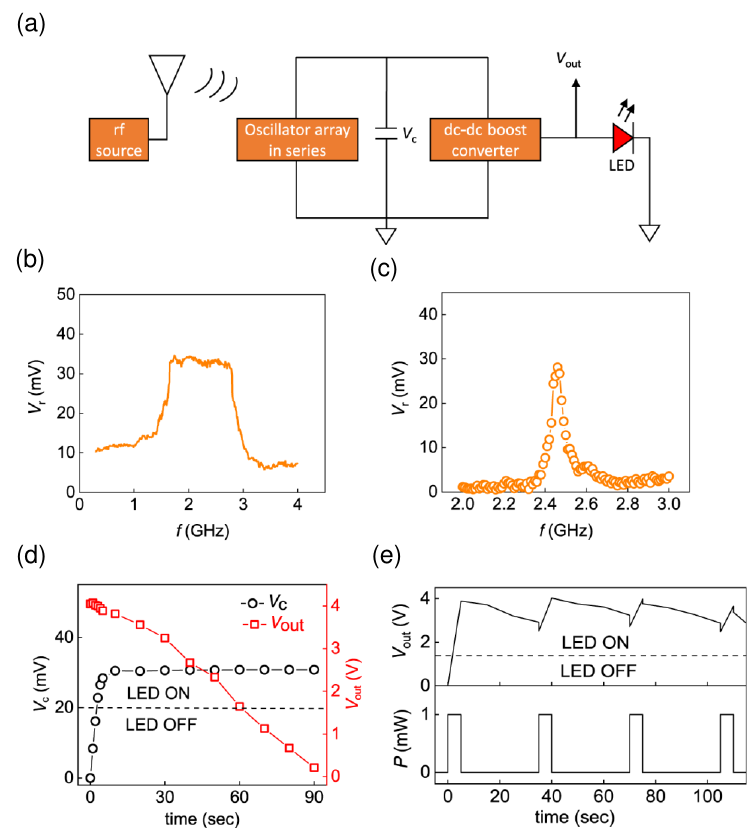}
    \caption{ (a) Schematic of the circuit used for energy harvesting using oscillator arrays. The RF source (a horn or a patch antenna) with a power of 0 dBm irradiates the array of oscillators that generates a DC voltage due to the spin diode effect. (b) and (c) DC voltage from 8 serially connected oscillators with horn antenna (6-7 dBi, 1-4 GHz) and patch antenna (0 dBm, 2.45 GHz), respectively. (d) Time dependence of voltage stored in the capacitor (left) and output voltage when the wireless output is off (right). (e) The output voltage from the circuit towards input 1 mW of power in patch antenna with an ON and OFF time of 5 and 30 s, respectively. Reprinted from~\cite{sharma2021electrically}}
    \label{fig:Energy}
\end{figure}

Spintronic oscillators can operate as microwave receivers ~\cite{fang2016giant,sidi2022size,bendjeddou2023electrical,bendjeddou2021radio,zhang2018ultrahigh,Kichin2023Rectification} using the spin-diode effect, in which these devices generate a DC signal in the presence of an RF input. Recently, Sharma\textit{ et al.,}~\cite{sharma2021electrically} demonstrated the potential of employing synchronized oscillator networks comprised of 8 serially connected MTJ STNOs to generate sufficient DC voltage from the untapped WiFi signal (2.4 GHz) \cite{sharma2021electrically}. The electrical setup for wireless energy harvesting is shown in Fig.~\ref{fig:Energy}a. Figure~\ref{fig:Energy}b and ~\ref{fig:Energy}c shows the rectified response from 8 STNOs with a wide-band horn antenna (0.3-4 GHz) and a microwave patch antenna (2.45 GHz), respectively. The 8 oscillators demonstrate a peak rectified voltage ranging from approximately 30 to 34 mV within the frequency span of 1.65--2.8 GHz when utilizing a wideband horn antenna, and approximately 30 mV when using a patch antenna at P$_{rf}$ = 0 dBm. The rectified voltage is then stored in a capacitor before dc-to-dc boost conversion to 3.5--4 V, which is higher than the voltage needed to light up an LED (1.6 V). The capacitor takes about 3-4 sec for charging, shown in Fig.~\ref{fig:Energy}d. This stored voltage is then up-converted to 4.1 V to light the LED for about one minute until the voltage drops below 1.6V (shown on the right axis of Fig.~\ref{fig:Energy}d). This demonstration highlights the application of spintronic oscillators as efficient energy harvesting from untapped microwave radiation and opens a pathway to utilize chip-based energy harvesting for wireless charging or signal detection.

Another proposed application of mutually synchronized oscillators is as highly sensitive magnetic field sensors. The frequency of spintronic oscillators strongly depends on the external magnetic field. This property can be effectively exploited to design STNO-based magnetic field sensors~\cite{albertsson2020magnetic,xie2023nanoscale}.

\begin{figure}[h!]
    \centering
    \includegraphics[width=1\linewidth]{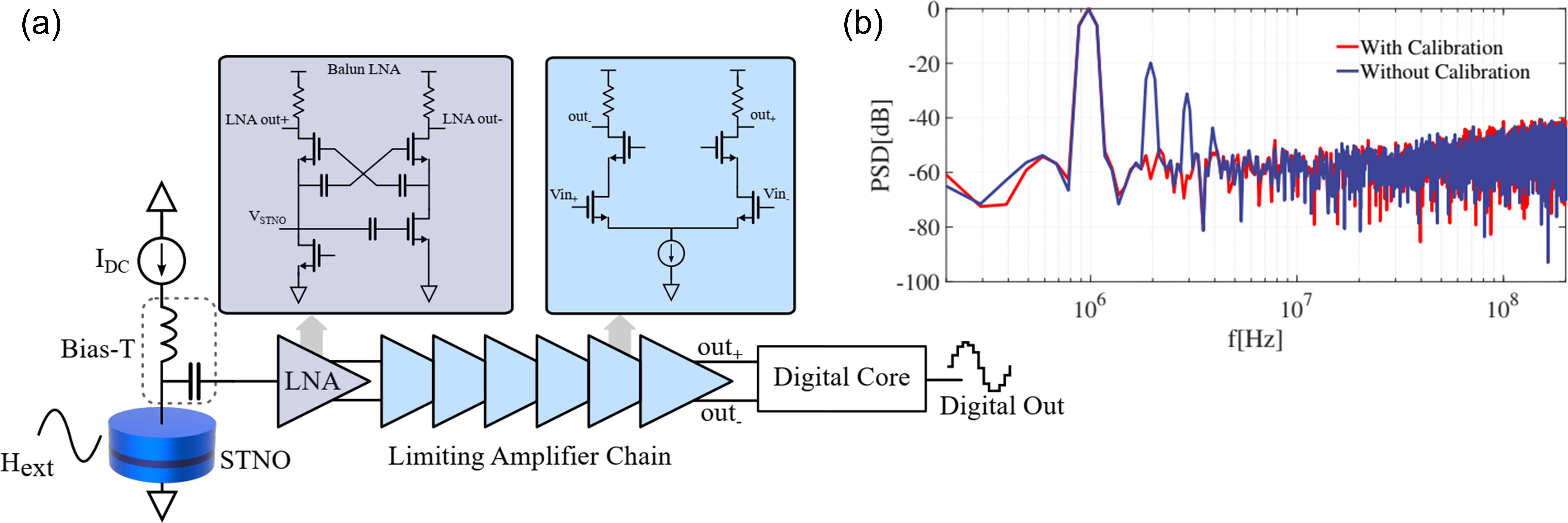}
    \caption{ (a) Schematic of a magnetic-to-digital converter (MDC) using an STNO. (b) PSD of the digital output of the STNO-based MDC with and without calibration. Reprinted with permission from~\cite{albertsson2020magnetic}}
    \label{fig:STNOM2D}
\end{figure}

In~\cite{albertsson2020magnetic}, an STNO is used as a magnetic-to-digital (MDC) converter, which is then connected to a classical frequency-to-digital converter (see Fig.~\ref{fig:STNOM2D}a,b). The proposed MDC has an SNR of 38 dB and a corresponding sensitivity of 5 Oe at a bandwidth of 10 MHz, which can be used for magnetic dynamics measurements at high rates and nanoscale dimensions. Similar to ADCs, MDCs can be implemented with mutually synchronized STNOs and SHNOs with significantly improved resolution and sensitivity. In ~\cite{xie2023nanoscale}, mutually synchronized SHNOs were used to improve the sensitivity of SHNO-based magnetic sensors (see Fig.~\ref{fig:STNOM2D}c,d,e) by almost an order of magnitude.

\section{Summary and Outlook}

In summary, the mutual synchronization of STNOs and SHNOs continues to hold great promise for both future fundamental studies and emerging applications. In this chapter, we have discussed different spintronic oscillator geometries, followed by their synchronization mechanisms. The experimental demonstration of mutual synchronization for various STNOs and SHNOs was discussed in detail. The benchmarking and evolution of phase noise in these oscillator networks have been presented. Demonstrated and suggested applications based on, or benefiting from, mutual synchronization are outlined. 

We believe that with improved fabrication techniques and a better understanding of their operation, mutual synchronization in STNOs and SHNOs can be extended to much larger oscillator arrays, including both one-dimensional (1D) and two-dimensional (2D) configurations, and potentially even into three-dimensional (3D) space. The substantial output power and low linewidth at high frequencies can be harnessed for various applications in wireless communication and signal processing. The ease of both series and parallel connections, coupled with individual control through different means, position these devices as promising candidates for unconventional computing. In the context of large-scale implementation, one bottleneck for computing applications lies in the one-to-one connectivity of oscillators in large arrays, which needs attention in the near future.

\medskip
\textbf{Acknowledgements} \par %delete if not applicable))
This work was partially supported by the Horizon 2020 research and innovation program No. 835068 "TOPSPIN". This work was also partially supported by the Swedish Research Council (VR Grant No. 2016-05980) and the Knut and Alice Wallenberg Foundation.

\bibliography{Bibliography}

\providecommand{\noopsort}[1]{}\providecommand{\singleletter}[1]{#1}%
\begin{thebibliography}{100}
\providecommand{\url}[1]{{#1}}
\providecommand{\urlprefix}{URL }
\expandafter\ifx\csname urlstyle\endcsname\relax
  \providecommand{\doi}[1]{DOI \discretionary{}{}{}#1}\else
  \providecommand{\doi}{DOI \discretionary{}{}{}\begingroup
  \urlstyle{rm}\Url}\fi

\bibitem{Slonczewski1996a}
J.~Slonczewski, J. Magn. Magn. Mater. \textbf{159}(1-2), L1 (1996)

\bibitem{berger1996emission}
L.~Berger, Phys.\ Rev.\ B \textbf{54}(13), 9353 (1996)

\bibitem{Baibich1988}
M.N. Baibich, J.M. Broto, A.~Fert, F.N. {Van Dau}, F.~Petroff, P.~Etienne,
  G.~Creuzet, A.~Friederich, J.~Chazelas, Phys. Rev. Lett. \textbf{61}(21),
  2472 (1988)

\bibitem{binasch1989enhanced}
G.~Binasch, P.~Gr{\"u}nberg, F.~Saurenbach, W.~Zinn, Phys.\ Rev.\ B
  \textbf{39}(7), 4828 (1989)

\bibitem{Miyazaki1995}
T.~Miyazaki, N.~Tezuka, J. Magn. Magn. Mater. \textbf{139}, L231 (1995)

\bibitem{Moodera1995}
J.S. Moodera, L.~Kinder, T.~Wong, R.~Meservey, Phys. Rev. Lett.
  \textbf{74}(16), 3273 (1995)

\bibitem{Katine2000}
J.A. Katine, F.J. Albert, R.A. Buhrman, E.B. Myers, D.C. Ralph, Phys. Rev.
  Lett. \textbf{84}, 3149 (2000)

\bibitem{Akerman2005sci}
J.~{\AA}kerman, Science \textbf{743}(9), 508 (2005)

\bibitem{Chen2016procieee}
T.~Chen, R.K. Dumas, A.~Eklund, P.K. Muduli, A.~Houshang, A.A. Awad,
  D\"{u}rrenfeld, B.G. Malm, A.~Rusu, J.~{\AA}kerman, P. IEEE \textbf{104}(10),
  1919 (2016)

\bibitem{Tsoi1998}
M.~Tsoi, A.G.M. Jansen, J.~Bass, W.C. Chiang, M.~Seck, V.~Tsoi, P.~Wyder, Phys.
  Rev. Lett. \textbf{80}, 4281 (1998)

\bibitem{Kiselev2003}
S.I. Kiselev, J.C. Sankey, I.N. Krivorotov, N.C. Emley, R.J. Schoelkopf, R.A.
  Buhrman, D.C. Ralph, Nature \textbf{425}(6956), 380 (2003)

\bibitem{Dyakonov1971jetp}
M.~D'yakonov, V.~Perel', Sov. Phys. JETP Lett. \textbf{13}(11), 467 (1971)

\bibitem{Hirsch1999}
J.E. Hirsch, Phys. Rev. Lett. \textbf{83}(9), 1834 (1999)

\bibitem{Liu2012prl}
L.~Liu, C.F. Pai, D.C. Ralph, R.A. Buhrman, Phys. Rev. Lett. \textbf{109}(18),
  186602 (2012)

\bibitem{liu2012science}
L.~Liu, C.F. Pai, Y.~Li, H.W. Tseng, D.C. Ralph, R.A. Buhrman, Science
  \textbf{336}, 555 (2012)

\bibitem{demidov2012nm}
V.E. Demidov, S.~Urazhdin, H.~Ulrichs, V.~Tiberkevich, A.~Slavin, D.~Baither,
  G.~Schmitz, S.O. Demokritov, Nat.\ Mater. \textbf{11}(12), 1028 (2012)

\bibitem{Kaka2005}
S.~Kaka, M.R. Pufall, W.H. Rippard, T.J. Silva, S.E. Russek, J.A. Katine,
  Nature \textbf{437}(7057), 389 (2005)

\bibitem{Mancoff2005}
F.B. Mancoff, N.D. Rizzo, B.N. Engel, S.~Tehrani, Nature \textbf{437}(7057),
  393 (2005)

\bibitem{Houshang2015natnano}
A.~Houshang, E.~Iacocca, P.~D{\"u}rrenfeld, S.R. Sani, J.~{\AA}kerman, R.K.
  Dumas, Nat.\ Nano. \textbf{11}(3), 280 (2016)

\bibitem{tsunegi2018scaling}
S.~Tsunegi, T.~Taniguchi, R.~Lebrun, K.~Yakushiji, V.~Cros, J.~Grollier,
  A.~Fukushima, S.~Yuasa, H.~Kubota, Sci. Rep. \textbf{8}(1), 1 (2018)

\bibitem{ralph2008jmmm}
D.C. Ralph, M.D. Stiles, J.\ Magn.\ Magn.\ Mater. \textbf{320}(7), 1190  (2008)

\bibitem{sinova2015revmodern}
J.~Sinova, S.O. Valenzuela, J.~Wunderlich, C.~Back, T.~Jungwirth, Rev.\ Mod.\
  Phys. \textbf{87}(4), 1213 (2015)

\bibitem{slavin2009nonlinear}
A.~Slavin, V.~Tiberkevich, IEEE Trans. Magn. \textbf{45}(4), 1875 (2009)

\bibitem{dvornik2018pra}
M.~Dvornik, A.A. Awad, J.~{\AA}kerman, Phys.\ Rev.\ Appl. \textbf{9}(1), 014017
  (2018)

\bibitem{Gilbert1955}
T.L. Gilbert, Phys. Rev. \textbf{100}, 1243 (1955)

\bibitem{rajabali2023injection}
M.~Rajabali, R.~Ovcharov, R.~Khymyn, H.~Fulara, A.~Kumar, A.~Litvinenko,
  M.~Zahedinejad, A.~Houshang, A.A. Awad, J.~{\AA}kerman, Phys.\ Rev.\ Appl.
  \textbf{19}(3), 034070 (2023)

\bibitem{sethi2023compensation}
P.~Sethi, D.~Sanz-Hern{\'a}ndez, F.~Godel, S.~Krishnia, F.~Ajejas, A.~Mizrahi,
  V.~Cros, D.~Markovi{\'c}, J.~Grollier, Phys.\ Rev.\ Appl. \textbf{19}(6),
  064018 (2023)

\bibitem{fulara2019spin}
H.~Fulara, M.~Zahedinejad, R.~Khymyn, A.~Awad, S.~Muralidhar, M.~Dvornik,
  J.~{\AA}kerman, Science Advances \textbf{5}(9), eaax8467 (2019)

\bibitem{bonetti2010experimental}
S.~Bonetti, V.~Tiberkevich, G.~Consolo, G.~Finocchio, P.~Muduli, F.~Mancoff,
  A.~Slavin, J.~{\AA}kerman, Phys.\ Rev.\ Lett. \textbf{105}(21), 217204 (2010)

\bibitem{Dumas2013prl}
R.K. Dumas, E.~Iacocca, S.~Bonetti, S.R. Sani, S.M. Mohseni, A.~Eklund,
  J.~Persson, O.~Heinonen, J.~\AA{}kerman, Phys. Rev. Lett. \textbf{110},
  257202 (2013)

\bibitem{jiang2018impact}
S.~Jiang, S.R. Etesami, S.~Chung, Q.T. Le, A.~Houshang, J.~{\AA}kerman, IEEE
  Magnetics Letters \textbf{9}, 1 (2018)

\bibitem{Tsoi2000}
M.~Tsoi, A.G.M. Jansen, J.~Bass, W.C. Chiang, V.~Tsoi, P.~Wyder, Nature
  \textbf{406}(6791), 46 (2000)

\bibitem{Sun1999}
J.~Sun, J. Magn. Magn. Mater. \textbf{202}(1), 157  (1999)

\bibitem{Myers1999}
E.B. Myers, D.C. Ralph, J.A. Katine, R.N. Louie, R.A. Buhrman, Science
  \textbf{285}(5429), 867 (1999)

\bibitem{Rippard2003}
W.H. Rippard, M.R. Pufall, T.J. Silva, Appl. Phys. Lett. \textbf{82}(8), 1260
  (2003)

\bibitem{Pufall2003}
M.R. Pufall, W.H. Rippard, T.J. Silva, Appl. Phys. Lett. \textbf{83}(2), 323
  (2003)

\bibitem{slavin2005prl}
A.~Slavin, V.~Tiberkevich, Phys. Rev. Lett. \textbf{95}, 237201 (2005)

\bibitem{bonetti2012prb}
S.~Bonetti, V.~Puliafito, G.~Consolo, V.S. Tiberkevich, A.N. Slavin,
  J.~\AA{}kerman, Phys. Rev. B \textbf{85}, 174427 (2012)

\bibitem{madami2011natnano}
M.~Madami, S.~Bonetti, G.~Consolo, S.~Tacchi, G.~Carlotti, G.~Gubbiotti, F.B.
  Mancoff, M.A. Yar, J.~Åkerman, Nature Nanotechnology \textbf{6}(10), 635
  (2011)

\bibitem{madami2015prb}
M.~Madami, E.~Iacocca, S.~Sani, G.~Gubbiotti, S.~Tacchi, R.K. Dumas,
  J.~\AA{}kerman, G.~Carlotti, Phys. Rev. B \textbf{92}, 024403 (2015)

\bibitem{mohseni2013science}
S.M. Mohseni, S.~Sani, J.~Persson, T.A. Nguyen, S.~Chung, Y.~Pogoryelov,
  P.~Muduli, E.~Iacocca, A.~Eklund, R.~Dumas, et~al., Science
  \textbf{339}(6125), 1295 (2013)

\bibitem{chung2016natcomm}
S.~Chung, A.~Eklund, E.~Iacocca, S.M. Mohseni, S.R. Sani, L.~Bookman, M.A.
  Hoefer, R.K. Dumas, J.~{\AA}kerman, Nat. Commun. \textbf{7}, 11209 (2016)

\bibitem{chung2018prl}
S.~Chung, Q.T. Le, M.~Ahlberg, A.A. Awad, M.~Weigand, I.~Bykova, R.~Khymyn,
  M.~Dvornik, H.~Mazraati, A.~Houshang, S.~Jiang, T.A. Nguyen, E.~Goering,
  G.~Sch{\"{u}}tz, J.~Gr{\"{a}}fe, J.~{\AA}kerman, Phys. Rev. Lett.
  \textbf{120}(21), 217204 (2018)

\bibitem{ahlberg2022natcomm}
M.~Ahlberg, S.~Chung, S.~Jiang, A.~Frisk, M.~Khademi, R.~Khymyn, A.A. Awad,
  Q.T. Le, H.~Mazraati, M.~Mohseni, M.~Weigand, I.~Bykova, F.~Gro{\ss},
  E.~Goering, G.~Sch\"{u}tz, J.~Gr\"{a}fe, J.~\AA{}kerman, Nat. Commun.
  \textbf{13}(1), 2462 (2022)

\bibitem{zhou2015natcomm}
Y.~Zhou, E.~Iacocca, A.~Awad, R.K. Dumas, F.C. Zhang, H.B. Braun, Nat. Commun.
  \textbf{6}, 8193 (2015)

\bibitem{pribiag2007magnetic}
V.~Pribiag, I.~Krivorotov, G.~Fuchs, P.~Braganca, O.~Ozatay, J.~Sankey,
  D.~Ralph, R.~Buhrman, Nat.\ Phys. \textbf{3}(7), 498 (2007)

\bibitem{mohseni2011pssrrl}
S.M. Mohseni, S.R. Sani, J.~Persson, T.N. {Anh Nguyen}, S.~Chung,
  Y.~Pogoryelov, J.~{\AA}kerman, Phys. status solidi - RRL. \textbf{5}(12), 432
  (2011)

\bibitem{jiang2023nl}
S.~Jiang, S.~Chung, Q.T. Le, P.K.J. Wong, W.~Zhang, J.~{\AA}kerman, Nano
  Letters \textbf{23}(4), 1159 (2023)

\bibitem{banuazizi2017order}
S.A.H. Banuazizi, S.R. Sani, A.~Eklund, M.M. Naiini, S.M. Mohseni, S.~Chung,
  P.~D{\"u}rrenfeld, B.G. Malm, J.~{\AA}kerman, Nanoscale \textbf{9}(5), 1896
  (2017)

\bibitem{Maehara2013large}
H.~{Maehara}, H.~{Kubota}, Y.~{Suzuki}, T.~{Seki}, K.~{Nishimura},
  Y.~{Nagamine}, K.~{Tsunekawa}, A.~{Fukushima}, A.M. {Deac}, K.~{Ando},
  S.~{Yuasa}, Appl. Phys. Exp. \textbf{6}(11), 113005 (2013)

\bibitem{maehara2014high}
H.~{Maehara}, H.~{Kubota}, Y.~{Suzuki}, T.~{Seki}, K.~{Nishimura},
  Y.~{Nagamine}, K.~{Tsunekawa}, A.~{Fukushima}, H.~{Arai}, T.~{Taniguchi},
  H.~{Imamura}, K.~{Ando}, S.~{Yuasa}, Appl. Phys. Exp. \textbf{7}(2), 023003
  (2014)

\bibitem{Houshang2018natcomm}
A.~Houshang, R.~Khymyn, H.~Fulara, A.~Gangwar, M.~Haidar, S.R. Etesami,
  R.~Ferreira, P.P. Freitas, M.~Dvornik, R.K. Dumas, J.~{\AA}kerman, Nat.\
  Commun. \textbf{9}(1), 4374 (2018)

\bibitem{litvinenko2022ultrafast}
A.~Litvinenko, A.~Sidi El~Valli, V.~Iurchuk, S.~Louis, V.~Tyberkevych,
  B.~Dieny, A.N. Slavin, U.~Ebels, Nano {L}ett. \textbf{22}(5), 1874 (2022)

\bibitem{Martin2013}
S.Y. Martin, C.~Thirion, C.~Hoarau, C.~Baraduc, B.~Di\'eny, Phys. Rev. B
  \textbf{88}, 024421 (2013)

\bibitem{zvezdin2022spin}
K.~Zvezdin, E.~Ekomasov, Physics of Metals and Metallography \textbf{123}(3),
  201 (2022)

\bibitem{ekomasov2021effect}
E.~Ekomasov, S.~Stepanov, K.~Zvezdin, N.~Pugach, G.~Antonov, Physics of Metals
  and Metallography \textbf{122}(3), 197 (2021)

\bibitem{Jenkins2016}
A.S. Jenkins, R.~Lebrun, E.~Grimaldi, S.~Tsunegi, P.~Bortolotti, H.~Kubota,
  K.~Yakushiji, A.~Fukushima, G.d. Loubens, O.~Klein, S.~Yuasa, V.~Cros, Nature
  Nanotech. \textbf{11}, 360 (2016)

\bibitem{tsunegi2016microwave}
S.~Tsunegi, K.~Yakushiji, A.~Fukushima, S.~Yuasa, H.~Kubota, Appl.\ Phys.\
  Lett. \textbf{109}(25), 252402 (2016)

\bibitem{Demidov2010}
V.E. Demidov, S.~Urazhdin, S.O. Demokritov, Nat.\ Mater. \textbf{9}(12), 984
  (2010)

\bibitem{Evarts2013}
E.R. Evarts, M.R. Pufall, W.H. Rippard, J. Appl. Phys. \textbf{113}(8), 083903
  (2013)

\bibitem{cai2023angular}
W.~Cai, A.~Kumar, A.~Du, K.~Shi, R.~Xiao, K.~Cao, J.~Yin, J.~{\AA}kerman,
  W.~Zhao, IEEE Electron Device Lett. \textbf{44}(5), 861 (2023)

\bibitem{Ranjbar2014}
M.~Ranjbar, P.~D\"{u}rrenfeld, M.~Haidar, E.~Iacocca, M.~Balinskiy, T.Q. Le,
  M.~Fazlali, A.~Houshang, A.A. Awad, R.K. Dumas, J.~{\AA}kerman, IEEE Magn.
  Lett. \textbf{5}, 3000504 (2014)

\bibitem{spicer2018time}
T.M. Spicer, P.S. Keatley, M.~Dvornik, T.H. Loughran, A.~Awad,
  P.~D{\"u}rrenfeld, A.~Houshang, M.~Ranjbar, J.~{\AA}kerman, V.V. Kruglyak,
  H.R. J., Appl.\ Phys.\ Lett. \textbf{113}(19) (2018)

\bibitem{Duan2014}
Z.~Duan, A.~Smith, L.~Yang, B.~Youngblood, J.~Lindner, V.E. Demidov, S.O.
  Demokritov, I.N. Krivorotov, Nat. Commun. \textbf{5}, 5616 (2014)

\bibitem{Demidov2014}
V.E. Demidov, S.~Urazhdin, A.~Zholud, A.V. Sadovnikov, S.O. Demokritov, Appl.
  Phys. Lett. \textbf{105}(17), 172410 (2014)

\bibitem{sato2019domain}
N.~Sato, K.~Schultheiss, L.~K{\"o}rber, N.~Puwenberg, T.~M{\"u}hl, A.~Awad,
  S.~Arekapudi, O.~Hellwig, J.~Fassbender, H.~Schultheiss, Phys.\ Rev.\ Lett.
  \textbf{123}(5), 057204 (2019)

\bibitem{durrenfeld2017nanoscale}
P.~D{\"u}rrenfeld, A.A. Awad, A.~Houshang, R.K. Dumas, J.~{\AA}kerman,
  Nanoscale \textbf{9}(3), 1285 (2017)

\bibitem{Mazraati2016apl}
H.~Mazraati, S.~Chung, A.~Houshang, M.~Dvornik, L.~Piazza, F.~Qejvanaj,
  S.~Jiang, T.Q. Le, J.~Weissenrieder, J.~{\AA}kerman, Appl. Phys. Lett.
  \textbf{109}(24), 242402 (2016)

\bibitem{xie2023nanoscale}
Y.~Xie, H.F.H. Cheung, G.D. Fuchs, arXiv preprint arXiv:2303.02478  (2023)

\bibitem{hache2020bipolar}
T.~Hache, Y.~Li, T.~Weinhold, B.~Scheumann, F.~Gon{\c{c}}alves, O.~Hellwig,
  J.~Fassbender, H.~Schultheiss, Appl.\ Phys.\ Lett. \textbf{116}(19) (2020)

\bibitem{zahedinejad2018cmos}
M.~Zahedinejad, H.~Mazraati, H.~Fulara, J.~Yue, S.~Jiang, A.~Awad,
  J.~{\AA}kerman, Appl.\ Phys.\ Lett. \textbf{112}(13), 132404 (2018)

\bibitem{kumar2022fabrication}
A.~Kumar, M.~Rajabali, V.H. Gonz{\'a}lez, M.~Zahedinejad, A.~Houshang,
  J.~{\AA}kerman, Nanoscale \textbf{14}, 1432 (2022)

\bibitem{behera2022energy}
N.~Behera, H.~Fulara, L.~Bainsla, A.~Kumar, M.~Zahedinejad, A.~Houshang,
  J.~{\AA}kerman, Phys.\ Rev.\ Appl. \textbf{18}(2), 024017 (2022)

\bibitem{choi2022voltage}
J.G. Choi, J.~Park, M.G. Kang, D.~Kim, J.S. Rieh, K.J. Lee, K.J. Kim, B.G.
  Park, Nat.\ Commun. \textbf{13}(1), 3783 (2022)

\bibitem{haidar2019natcomm}
M.~Haidar, A.A. Awad, M.~Dvornik, R.~Khymyn, A.~Houshang, J.~{\AA}kerman,
  Nature Communications \textbf{10}(1), 2362 (2019)

\bibitem{awad2020apl}
A.A. Awad, A.~Houshang, M.~Zahedinejad, R.~Khymyn, J.~{\AA}kerman, Appl.\
  Phys.\ Lett. \textbf{116}(23), 232401 (2020)

\bibitem{Rippard2005}
W.H. Rippard, M.R. Pufall, S.~Kaka, T.J. Silva, S.E. Russek, J.A. Katine, Phys.
  Rev. Lett. \textbf{95}, 067203 (2005)

\bibitem{zhou2008apl}
Y.~{Zhou}, J.~{Persson}, S.~{Bonetti}, J.~{{\AA}kerman}, Appl.\ Phys.\ Lett.
  \textbf{92}(9), 092505 (2008)

\bibitem{Georges2008}
B.~Georges, J.~Grollier, M.~Darques, V.~Cros, C.~Deranlot, B.~Marcilhac,
  G.~Faini, A.~Fert, Phys. Rev. Lett. \textbf{101}, 017201 (2008)

\bibitem{romera2016apl}
M.~Romera, P.~Talatchian, R.~Lebrun, K.J. Merazzo, P.~Bortolotti, L.~Vila, J.D.
  Costa, R.~Ferreira, P.P. Freitas, M.C. Cyrille, U.~Ebels, V.~Cros,
  J.~Grollier, Appl.\ Phys.\ Lett. \textbf{109}(25), 252404 (2016)

\bibitem{jue2018apl}
E.~Jué, M.R. Pufall, W.H. Rippard, Appl.\ Phys.\ Lett. \textbf{112}(10),
  102403 (2018)

\bibitem{Tortarolo2018scirep}
M.~Tortarolo, B.~Lacoste, J.~Hem, C.~Dieudonn{\'{e}}, M.C. Cyrille, J.A.
  Katine, D.~Mauri, A.~Zeltser, L.D. Buda-Prejbeanu, U.~Ebels, Sci. Rep.
  \textbf{8}(1), 1 (2018)

\bibitem{hem2019prb}
J.~Hem, L.D. Buda-Prejbeanu, U.~Ebels, Phys. Rev. B \textbf{100}, 054414 (2019)

\bibitem{letang2019prb}
J.~L\'etang, S.~Petit-Watelot, M.W. Yoo, T.~Devolder, K.~Bouzehouane, V.~Cros,
  J.V. Kim, Phys. Rev. B \textbf{100}, 144414 (2019)

\bibitem{Hache2019APL}
T.~Hache, T.~Weinhold, K.~Schultheiss, J.~Stigloher, F.~Vilsmeier, C.~Back,
  S.~Arekapudi, O.~Hellwig, J.~Fassbender, H.~Schultheiss, Appl.\ Phys.\ Lett.
  \textbf{114}(10), 102403 (2019)

\bibitem{urazhdin2010fractional}
S.~Urazhdin, P.~Tabor, V.~Tiberkevich, A.~Slavin, Phys.\ Rev.\ Lett.
  \textbf{105}(10), 104101 (2010)

\bibitem{zhou2007jap}
Y.~{Zhou}, J.~{Persson}, J.~{{\AA}kerman}, J.\ Appl.\ Phys. \textbf{101}(9),
  09A510 (2007)

\bibitem{singh2017integer}
H.~Singh, K.~Konishi, S.~Bhuktare, A.~Bose, S.~Miwa, A.~Fukushima,
  K.~Yakushiji, S.~Yuasa, H.~Kubota, Y.~Suzuki, et~al., Phys.\ Rev.\ Appl.
  \textbf{8}(6), 064011 (2017)

\bibitem{litvinenko2021analog}
A.~Litvinenko, P.~Sethi, C.~Murapaka, A.~Jenkins, V.~Cros, P.~Bortolotti,
  R.~Ferreira, B.~Dieny, U.~Ebels, Phys.\ Rev.\ Appl. \textbf{16}(2), 024048
  (2021)

\bibitem{romera2018nt}
M.~Romera, P.~Talatchian, S.~Tsunegi, F.A. Araujo, V.~Cros, P.~Bortolotti,
  J.~Trastoy, K.~Yakushiji, A.~Fukushima, H.~Kubota, et~al., Nature
  \textbf{563}, 230 (2018)

\bibitem{zahedinejad2020nt}
M.~Zahedinejad, A.A. Awad, S.~Muralidhar, R.~Khymyn, H.~Fulara, H.~Mazraati,
  M.~Dvornik, J.~{\AA}kerman, Nat.\ Nano. \textbf{15}(1), 47 (2020)

\bibitem{mcgoldrick2022ising}
B.C. McGoldrick, J.Z. Sun, L.~Liu, Phys.\ Rev.\ Appl. \textbf{17}(1), 014006
  (2022)

\bibitem{finocchio2023roadmap}
G.~Finocchio, S.~Bandyopadhyay, P.~Lin, G.~Pan, J.J. Yang, R.~Tomasello,
  C.~Panagopoulos, M.~Carpentieri, V.~Puliafito, J.~{\AA}kerman, et~al., Nano
  Futures \textbf{8}, 012001 (2024)

\bibitem{chumak2022advances}
A.V. Chumak, P.~Kabos, M.~Wu, C.~Abert, C.~Adelmann, A.~Adeyeye,
  J.~{\AA}kerman, F.G. Aliev, A.~Anane, A.~Awad, et~al., IEEE Trans. Magn.
  \textbf{58}(6), 1 (2022)

\bibitem{Slavin2006}
A.N. Slavin, V.S. Tiberkevich, Phys. Rev. B \textbf{74}, 104401 (2006)

\bibitem{slavin2005nonlinear}
A.~Slavin, V.~Tiberkevich, Physical Review B \textbf{72}(9), 092407 (2005)

\bibitem{Demidov2014b}
V.E. Demidov, H.~Ulrichs, S.V. Gurevich, S.O. Demokritov, V.S. Tiberkevich,
  A.N. Slavin, A.~Zholud, S.~Urazhdin, Nat. Commun. \textbf{5}, 3179 (2014)

\bibitem{Lebrun2015}
R.~Lebrun, A.~Jenkins, A.~Dussaux, N.~Locatelli, S.~Tsunegi, E.~Grimaldi,
  H.~Kubota, P.~Bortolotti, K.~Yakushiji, J.~Grollier, A.~Fukushima, S.~Yuasa,
  V.~Cros, Phys. Rev. Lett. \textbf{115}, 017201 (2015)

\bibitem{Kendziorczyk2014}
T.~Kendziorczyk, S.O. Demokritov, T.~Kuhn, Phys. Rev. B \textbf{90}, 054414
  (2014)

\bibitem{Kendziorczyk2016prb}
T.~Kendziorczyk, T.~Kuhn, Phys. Rev. B \textbf{93}, 134413 (2016)

\bibitem{Pufall2006}
M.R. Pufall, W.H. Rippard, S.E. Russek, S.~Kaka, J.A. Katine, Phys. Rev. Lett.
  \textbf{97}, 087206 (2006)

\bibitem{Erokhin2014}
S.~Erokhin, D.~Berkov, Phys. Rev. B \textbf{89}, 144421 (2014)

\bibitem{flovik2016describing}
V.~Flovik, F.~Macia, E.~Wahlstr{\"o}m, Sci. Rep. \textbf{6}(1), 32528 (2016)

\bibitem{awad2017natphys}
A.A. Awad, P.~D{\"{u}}rrenfeld, A.~Houshang, M.~Dvornik, E.~Iacocca, R.K.
  Dumas, J.~{\AA}kerman, Nat. Phys. \textbf{13}(November), 292 (2017)

\bibitem{RuotoloA.2009}
A.~Ruotolo, V.~Cros, B.~Georges, A.~Dussaux, J.~Grollier, C.~Deranlot,
  R.~Guillemet, K.~Bouzehouane, S.~Fusil, A.~Fert, Nature Nanotech.
  \textbf{4}(8), 528 (2009)

\bibitem{sharma2021electrically}
R.~Sharma, R.~Mishra, T.~Ngo, Y.X. Guo, S.~Fukami, H.~Sato, H.~Ohno, H.~Yang,
  Nat.\ Commun. \textbf{12}(1), 1 (2021)

\bibitem{Sani2013ntc}
S.~Sani, J.~Persson, S.M. Mohseni, Y.~Pogoryelov, P.K. Muduli, A.~Eklund,
  G.~Malm, M.~K\"{a}ll, A.~Dmitriev, J.~\AA{}kerman, Nat. Commun. \textbf{4},
  2731 (2013)

\bibitem{pogoryelov2011b}
Y.~{Pogoryelov}, P.K. {Muduli}, S.~{Bonetti}, E.~{Iacocca}, F.~{Mancoff},
  J.~{{\AA}kerman}, Appl. Phys. Lett. \textbf{98}(19), 192501 (2011)

\bibitem{pogoryelov2012combined}
Y.~Pogoryelov, P.~Muduli, J.~Akerman, IEEE Trans. Magn. \textbf{48}(11), 4077
  (2012)

\bibitem{castro2022mutual}
M.~Castro, D.~Mancilla-Almonacid, B.~Dieny, S.~Allende, L.~Buda-Prejbeanu,
  U.~Ebels, Sci. Rep. \textbf{12}(1), 12030 (2022)

\bibitem{Locatelli2015}
N.~Locatelli, A.~Hamadeh, F.~Abreu~Araujo, A.D. Belanovsky, P.N. Skirdkov,
  R.~Lebrun, V.V. Naletov, K.A. Zvezdin, M.~Mu{\~n}oz, J.~Grollier, O.~Klein,
  V.~Cros, G.~de~Loubens, Sci. Rep. \textbf{5}, 17039 (2015)

\bibitem{abreu2016controlling}
F.~Abreu~Araujo, J.~Grollier, J.\ Appl.\ Phys. \textbf{120}(10) (2016)

\bibitem{lebrun2017mutual}
R.~Lebrun, S.~Tsunegi, P.~Bortolotti, H.~Kubota, A.~Jenkins, M.~Romera,
  K.~Yakushiji, A.~Fukushima, J.~Grollier, S.~Yuasa, et~al., Nat.\ Commun.
  \textbf{8}(1), 15825 (2017)

\bibitem{romera2022binding}
M.~Romera, P.~Talatchian, S.~Tsunegi, K.~Yakushiji, A.~Fukushima, H.~Kubota,
  S.~Yuasa, V.~Cros, P.~Bortolotti, M.~Ernoult, et~al., Nat.\ Commun.
  \textbf{13}(1), 883 (2022)

\bibitem{kumar2023robust}
A.~Kumar, H.~Fulara, R.~Khymyn, A.~Litvinenko, M.~Zahedinejad, M.~Rajabali,
  X.~Zhao, N.~Behera, A.~Houshang, A.A. Awad, J.~{\AA}kerman, Nano Lett.
  \textbf{23}, 6720 (2023)

\bibitem{behera2023ultra}
N.~Behera, A.K. Chaurasiya, V.H. Gonz{\'a}lez, A.~Litvinenko, L.~Bainsla,
  A.~Kumar, R.~Khymyn, A.A. Awad, H.~Fulara, J.~{\AA}kerman, Advanced Materials
  \textbf{36}(5), 2305002 (2024)

\bibitem{Zahedinejad2022natmat}
M.~Zahedinejad, H.~Fulara, R.~Khymyn, A.~Houshang, M.~Dvornik, S.~Fukami,
  S.~Kanai, H.~Ohno, J.~{\AA}kerman, Nat.\ Mater. \textbf{21}(1), 81 (2022)

\bibitem{gonzalez2022apl}
V.H. González, R.~Khymyn, H.~Fulara, A.A. Awad, J.~Åkerman, Appl.\ Phys.\
  Lett. \textbf{121}(25), 252404 (2022)

\bibitem{mohammad2016nanotech}
B.~Mohammad, M.A. Jaoude, V.~Kumar, D.M. Al~Homouz, H.A. Nahla, M.~Al-Qutayri,
  N.~Christoforou, Nanotechnology Reviews \textbf{5}(3), 311 (2016)

\bibitem{Khademi2023Large}
M.~Khademi, A.~Kumar, M.~Rajabali, S.P. Dash, J.~Åkerman, IEEE Electron Device
  Lett. \textbf{45}(2), 268  (2024)

\bibitem{muralidhar2022optothermal}
S.~Muralidhar, A.~Houshang, A.~Alem{\'a}n, R.~Khymyn, A.A. Awad,
  J.~{\AA}kerman, Appl.\ Phys.\ Lett. \textbf{120}(26) (2022)

\bibitem{moradi2019spin}
F.~Moradi, H.~Farkhani, B.~Zeinali, H.~Ghanatian, J.M.A. Pelloux-Prayer,
  T.~Boehnert, M.~Zahedinejad, H.~Heidari, V.~Nabaei, R.~Ferreira, et~al.,
  arXiv preprint arXiv:1912.01347  (2019)

\bibitem{houssameddine2008spin}
D.~Houssameddine, S.~Florez, J.~Katine, J.P. Michel, U.~Ebels, D.~Mauri,
  O.~Ozatay, B.~Delaet, B.~Viala, L.~Folks, B.D. Terris, M.C. Cyrille, Appl.\
  Phys.\ Lett. \textbf{93}(2), 022505 (2008)

\bibitem{kubota2013spin}
H.~Kubota, K.~Yakushiji, A.~Fukushima, S.~Tamaru, M.~Konoto, T.~Nozaki,
  S.~Ishibashi, T.~Saruya, S.~Yuasa, T.~Taniguchi, H.~Arai, H.~Imamura, Appl.
  Phys. Exp. \textbf{6}(10), 103003 (2013)

\bibitem{zeng2012high}
Z.~Zeng, P.K. Amiri, I.N. Krivorotov, H.~Zhao, G.~Finocchio, J.P. Wang, J.A.
  Katine, Y.~Huai, J.~Langer, K.~Galatsis, K.L. Wang, H.~Jiang, ACS {N}ano
  \textbf{6}(7), 6115 (2012)

\bibitem{deac2008bias}
A.M. Deac, A.~Fukushima, H.~Kubota, H.~Maehara, Y.~Suzuki, S.~Yuasa,
  Y.~Nagamine, K.~Tsunekawa, D.D. Djayaprawira, N.~Watanabe, Nat.\ Phys.
  \textbf{4}(10), 803 (2008)

\bibitem{zeng2011enhancement}
Z.~Zeng, P.~Upadhyaya, P.~Khalili~Amiri, K.~Cheung, J.~Katine, J.~Langer,
  K.~Wang, H.~Jiang, Appl.\ Phys.\ Lett. \textbf{99}(3), 032503 (2011)

\bibitem{costa2017high}
J.D. {Costa}, S.~{Serrano-Guisan}, B.~{Lacoste}, A.S. {Jenkins},
  T.~{B{\"o}hnert}, M.~{Tarequzzaman}, J.~{Borme}, F.L. {Deepak}, E.~{Paz},
  J.~{Ventura}, R.~{Ferreira}, P.P. {Freitas}, Sci. Rep. \textbf{7}(1), 1
  (2017)

\bibitem{seki2014high}
T.~Seki, Y.~Sakuraba, H.~Arai, M.~Ueda, R.~Okura, H.~Imamura, K.~Takanashi,
  Appl.\ Phys.\ Lett. \textbf{105}(9), 092406 (2014)

\bibitem{tsunegi2014high}
S.~{Tsunegi}, H.~{Kubota}, K.~{Yakushiji}, M.~{Konoto}, S.~{Tamaru},
  A.~{Fukushima}, H.~{Arai}, H.~{Imamura}, E.~{Grimaldi}, R.~{Lebrun},
  J.~{Grollier}, V.~{Cros}, S.~{Yuasa}, Appl. Phys. Exp. \textbf{7}(6), 063009
  (2014)

\bibitem{dussaux2014large}
A.~{Dussaux}, E.~{Grimaldi}, B.~{Rache Salles}, A.S. {Jenkins}, A.V.
  {Khvalkovskiy}, P.~{Bortolotti}, J.~{Grollier}, H.~{Kubota}, A.~{Fukushima},
  K.~{Yakushiji}, S.~{Yuasa}, V.~{Cros}, A.~{Fert}, Appl.\ Phys.\ Lett.
  \textbf{105}(2), 022404 (2014)

\bibitem{dussaux2010large}
A.~{Dussaux}, B.~{Georges}, J.~{Grollier}, V.~{Cros}, A.V. {Khvalkovskiy},
  A.~{Fukushima}, M.~{Konoto}, H.~{Kubota}, K.~{Yakushiji}, S.~{Yuasa}, K.A.
  {Zvezdin}, K.~{Ando}, A.~{Fert}, Nat.\ Commun. \textbf{1}(1), 1 (2010)

\bibitem{rippard2004}
W.H. Rippard, M.R. Pufall, S.~Kaka, S.E. Russek, T.J. Silva, Phys. Rev. Lett.
  \textbf{92}, 027201 (2004)

\bibitem{sani2013mutually}
S.~Sani, J.~Persson, S.M. Mohseni, Y.~Pogoryelov, P.~Muduli, A.~Eklund,
  G.~Malm, M.~K{\"a}ll, A.~Dmitriev, J.~{\AA}kerman, Nat.\ Commun.
  \textbf{4}(1), 1 (2013)

\bibitem{chen2020spin}
J.R. Chen, A.~Smith, E.A. Montoya, J.G. Lu, I.N. Krivorotov, Commun. Phys.
  \textbf{3}(1), 1 (2020)

\bibitem{litvinenko2023phase}
A.~Litvinenko, A.~Kumar, M.~Rajabali, A.A. Awad, R.~Khymyn, J.~{\AA}kerman,
  Appl.\ Phys.\ Lett. \textbf{122}(22) (2023)

\bibitem{wittrock2021stabilization}
S.~Wittrock, M.~Krei{\ss}ig, B.~Lacoste, A.~Litvinenko, P.~Talatchian,
  F.~Protze, F.~Ellinger, R.~Ferreira, R.~Lebrun, P.~Bortolotti, et~al., arXiv
  preprint arXiv:2110.13073  (2021)

\bibitem{kim2008prl}
J.V. Kim, V.~Tiberkevich, A.N. Slavin, Phys. Rev. Lett. \textbf{100}, 017207
  (2008)

\bibitem{jiang2020reduced}
S.~Jiang, R.~Khymyn, S.~Chung, T.Q. Le, L.H. Diez, A.~Houshang, M.~Zahedinejad,
  D.~Ravelosona, J.~{\AA}kerman, Appl.\ Phys.\ Lett. \textbf{116}(7) (2020)

\bibitem{stamps2014jphysd}
R.L. Stamps, S.~Breitkreutz, J.~Åkerman, A.V. Chumak, Y.~Otani, G.E.W. Bauer,
  J.U. Thiele, M.~Bowen, S.A. Majetich, M.~Kläui, I.L. Prejbeanu, B.~Dieny,
  N.M. Dempsey, B.~Hillebrands, J. Phys. D \textbf{47}(33), 333001 (2014)

\bibitem{dieny2020natelectron}
B.~Dieny, I.L. Prejbeanu, K.~Garello, P.~Gambardella, P.~Freitas, R.~Lehndorff,
  W.~Raberg, U.~Ebels, S.O. Demokritov, J.~Akerman, A.~Deac, P.~Pirro,
  C.~Adelmann, A.~Anane, A.V. Chumak, A.~Hirohata, S.~Mangin, S.O. Valenzuela,
  M.C. Onba{\c{s}}l{\i}, M.~d'Aquino, G.~Prenat, G.~Finocchio, L.~Lopez-Diaz,
  R.~Chantrell, O.~Chubykalo-Fesenko, P.~Bortolotti, Nat. Electron.
  \textbf{3}(8), 446 (2020)

\bibitem{choi2014spin}
H.S. Choi, S.Y. Kang, S.J. Cho, I.Y. Oh, M.~Shin, H.~Park, C.~Jang, B.C. Min,
  S.I. Kim, S.Y. Park, et~al., Sci. Rep. \textbf{4}(1), 5486 (2014)

\bibitem{manfrini2009apl}
M.~{Manfrini}, T.~{Devolder}, J.V. {Kim}, P.~{Crozat}, N.~{Zerounian},
  C.~{Chappert}, W.~{van Roy}, L.~{Lagae}, G.~{Hrkac}, T.~{Schrefl}, Appl.
  Phys. Lett. \textbf{95}(19), 192507 (2009)

\bibitem{manfrini2011jap}
M.~{Manfrini}, T.~{Devolder}, J.V. {Kim}, P.~{Crozat}, C.~{Chappert}, W.~{van
  Roy}, L.~{Lagae}, J. Appl. Phys. \textbf{109}(8), 083940 (2011)

\bibitem{pufall2005}
M.R. {Pufall}, W.H. {Rippard}, S.~{Kaka}, T.J. {Silva}, S.E. {Russek}, Appl.
  Phys. Lett. \textbf{86}(8), 082506 (2005)

\bibitem{muduli2010}
P.K. {Muduli}, Y.~{Pogoryelov}, S.~{Bonetti}, G.~{Consolo}, F.~{Mancoff},
  J.~{{\AA}kerman}, Phys. Rev. B \textbf{81}(14), 140408 (2010)

\bibitem{pogoryelov2011a}
Y.~Pogoryelov, P.K. Muduli, S.~Bonetti, F.~Mancoff, J.~{\AA}kerman, Appl. Phys.
  Lett. \textbf{98}(19), 192506 (2011)

\bibitem{Choi2014scirep}
H.S. Choi, S.Y. Kang, S.J. Cho, I.Y. Oh, M.~Shin, H.~Park, C.~Jang, B.C. Min,
  S.I. Kim, S.Y. Park, C.S. Park, Sci. Rep. \textbf{4}, 1 (2014)

\bibitem{Sharma2014apl}
R.~Sharma, P.~D\"urrenfeld, E.~Iacocca, O.G. Heinonen, J.~\AA{}kerman, P.K.
  Muduli, Appl. Phys. Lett. \textbf{105}(13), 132404 (2014)

\bibitem{litvinenko2020ultrafast}
A.~Litvinenko, V.~Iurchuk, P.~Sethi, S.~Louis, V.~Tyberkevych, J.~Li,
  A.~Jenkins, R.~Ferreira, B.~Dieny, A.~Slavin, et~al., Nano {L}ett.
  \textbf{20}(8), 6104 (2020)

\bibitem{csaba2020coupled}
G.~Csaba, W.~Porod, Appl. Phys. Rev. \textbf{7}(1), 011302 (2020)

\bibitem{torrejon2017nt}
J.~Torrejon, M.~Riou, F.A. Araujo, S.~Tsunegi, G.~Khalsa, D.~Querlioz,
  P.~Bortolotti, V.~Cros, K.~Yakushiji, A.~Fukushima, H.~Kubota, S.~Yuasa, M.D.
  Stiles, J.~Grollier, Nature \textbf{547}(7664), 428 (2017)

\bibitem{albertsson2021ultrafast}
D.I. Albertsson, M.~Zahedinejad, A.~Houshang, R.~Khymyn, J.~{\AA}kerman,
  A.~Rusu, Appl.\ Phys.\ Lett. \textbf{118}(11) (2021)

\bibitem{houshang2022prappl}
A.~Houshang, M.~Zahedinejad, S.~Muralidhar, J.~Ch{\c{e}}ci{\'{n}}ski,
  R.~Khymyn, M.~Rajabali, H.~Fulara, A.A. Awad, M.~Dvornik, J.~{\AA}kerman,
  Phys.\ Rev.\ Appl. \textbf{17}(1), 014003 (2022)

\bibitem{litvinenko2023spinwave}
A.~Litvinenko, R.~Khymyn, V.H. Gonz{\'a}lez, R.~Ovcharov, A.A. Awad,
  V.~Tyberkevych, A.~Slavin, J.~{\AA}kerman, Commun. Phys. \textbf{6}(1), 227
  (2023)

\bibitem{gonzalez2024spintronic}
V.H. Gonz{\'a}lez, A.~Litvinenko, A.~Kumar, R.~Khymyn, J.~{\AA}kerman, arXiv
  preprint arXiv:2403.13564  (2024)

\bibitem{kumar2024spin}
A.~Kumar, V.H. Gonz{\'a}lez, N.~Behera, R.~Khymyn, A.A. Awad, J.~{\AA}kerman,
  et~al., arXiv preprint arXiv:2402.00586  (2024)

\bibitem{fang2016giant}
B.~Fang, M.~Carpentieri, X.~Hao, H.~Jiang, J.A. Katine, I.N. Krivorotov,
  B.~Ocker, J.~Langer, K.L. Wang, B.~Zhang, et~al., Nat.\ Commun.
  \textbf{7}(1), 11259 (2016)

\bibitem{sidi2022size}
A.~Sidi El~Valli, V.~Iurchuk, G.~Lezier, I.~Bendjeddou, R.~Lebrun, N.~Lamard,
  A.~Litvinenko, J.~Langer, J.~Wrona, L.~Vila, et~al., Appl.\ Phys.\ Lett.
  \textbf{120}(1) (2022)

\bibitem{bendjeddou2023electrical}
I.~Bendjeddou, M.J. Garcia, A.S. El~Valli, A.~Litvinenko, V.~Cros, U.~Ebels,
  A.~Jenkins, R.~Ferreira, R.~Dutra, D.~Morche, et~al., IEEE Trans. Microw.
  Theory Tech.  (2023)

\bibitem{bendjeddou2021radio}
A.S. El~Valli, et~al., in \emph{2021 19th IEEE International New Circuits and
  Systems Conference (NEWCAS)} (IEEE, 2021), pp. 1--4

\bibitem{zhang2018ultrahigh}
L.~Zhang, B.~Fang, J.~Cai, M.~Carpentieri, V.~Puliafito, F.~Garesc{\`\i}, P.K.
  Amiri, G.~Finocchio, Z.~Zeng, Appl.\ Phys.\ Lett. \textbf{113}(10) (2018)

\bibitem{Kichin2023Rectification}
G.~Kichin, P.~Skirdkov, K.~Zvezdin, Phys. Rev. Appl. \textbf{20}, 044078 (2023)

\bibitem{albertsson2020magnetic}
D.I. Albertsson, J.~{\AA}kerman, A.~Rusu, IEEE Trans. Nano. \textbf{19}, 565
  (2020)

\end{thebibliography}
\end{document}